\documentclass{aa}  

\usepackage[switch]{lineno} 

\usepackage{graphicx}
\usepackage{txfonts}
\usepackage{multirow}
\usepackage{colortbl}
\usepackage{dirtytalk}
\usepackage{amssymb}
\usepackage[normalem]{ulem}
\usepackage[colorlinks=True,citecolor=blue]{hyperref}
\usepackage{cleveref}
\crefformat{section}{\S#2#1#3} 
\crefformat{subsection}{\S#2#1#3}
\crefformat{subsubsection}{\S#2#1#3}
\usepackage{tikz}

\makeatletter
\renewcommand*\aa@pageof{, page \thepage{} of \pageref*{LastPage}}
\makeatother

\begin{document} 

   \title{Chasing ICM cooling and AGN feedback from the macro to the meso scales in the galaxy cluster ZwCl~235}
  
\titlerunning{Chasing ICM cooling and AGN feedback in ZwCl~235}
   \author{F. Ubertosi
          \inst{1,2}
          \and
          M. Gitti\inst{1,3}
          \and
          F. Brighenti\inst{1,4}
          }

   \institute{Dipartimento di Fisica e Astronomia, Università di Bologna, via Gobetti 93/2, I-40129 Bologna, Italy\\
              \email{francesco.ubertosi2@unibo.it} \\
    Istituto Nazionale di Astrofisica (INAF) - Osservatorio di Astrofisica e Scienza dello Spazio (OAS), via Gobetti 101, I-40129 Bologna, Italy\\
    Istituto Nazionale di Astrofisica (INAF) - Istituto di Radioastronomia (IRA), via Gobetti 101, I-40129 Bologna, Italy\\
    University of California Observatories/Lick Observatory, Department of Astronomy and Astrophysics, University of California, Santa Cruz, CA 95064, USA}
              
   \date{Accepted for publication in A\&A}

 
  \abstract
  {}
   {We aim to investigate the AGN/ICM interplay in ZwCl~235, a galaxy cluster with high X-ray flux, an extended central radio galaxy, and evidence of multi-phase gas at its center.}
   {Using archival data from the \textit{Chandra} telescope, the VLASS survey, the LOTSS survey and the VLBA telescope, we perform a complete study of ZwCl~235, dissecting the dynamics of the ICM, the thermodynamic state of the central gas, and the properties of the BCG. By means of radial profiles and 2D spectral maps, we measure the temperature, entropy and cooling time of the ICM, and we compare the morphology of the central radio galaxy with the surrounding medium.}
   {We find evidence that ZwCl~235 is a sloshing cool core cluster in which the activity of the central galaxy has excavated a pair of cavities and possibly uplifted enriched gas to an altitude of $\sim$30 kpc. In the cluster core, the lowest entropy ICM is preferentially found in a 20 kpc-long filament tangential to the southern radio lobe of the AGN. We argue that the observed cool ($\sim$1.3 keV) filament is likely produced by a combination of sloshing and stimulated ICM cooling, that may be fueling the central supermassive black hole. Additionally, we determine that the X-ray emission of the BCG originates from a $\sim$1.4 keV plasma kernel which extends for 5 kpc in radius and has a short cooling time ($\sim$240 Myr), and could represent the thermal corona of the BCG.
   }
   {Overall, we propose that several sources (the large scale ICM, the low entropy filament and the $\sim$1.4 keV kernel) of cold material are currently feeding the central AGN, and that the ICM cooling cycle expectations are met from the macro scales (between 5 - 100 kpc) to the meso scales ($\leq$5 kpc) of AGN feedback.}

   \keywords{galaxies: clusters: intracluster medium --
                X-rays: galaxies: clusters --
                radio continuum: galaxies -- galaxies: active 
               }

   \maketitle
%

\section{Introduction}
\label{intro}
The cores of galaxy clusters represent the region where the interplay between different cluster components - namely the intracluster medium (ICM), the brightest cluster galaxy (BCG) and the central active galactic nucleus (AGN) - is manifested (e.g., for reviews \citealt{2007ARA&A..45..117M,2012NJPh...14e5023M,2012AdAst2012E...6G,2017FrASS...4...42M,2017FrASS...4...10C,2020NatAs...4...10G,2021Univ....7..142E}). In particular, the investigation of \say{cool core} galaxy clusters (see e.g., \citealt{2001ApJ...560..194M,2010A&A...513A..37H,2014A&A...572A..46B} for a discussion of cool core systems) with a multi-wavelength approach has recently allowed to shed new lights on how AGN accretion and ejection mechanisms are coupled with the thermodynamic properties of the surrounding ICM (for recent works see e.g., \citealt{2019MNRAS.489..349R,2019MNRAS.485..229R,2019ApJ...870...57V,2021A&A...649A..23C,2022MNRAS.510.2327M,2022NatAs...6..109M,2022arXiv220109330T}). 
\\ \indent The high resolution \textit{Chandra} and \textit{XMM-Newton} X-ray observations of these objects have led to the measurements of the central gas temperature and density, finding that cooling of the ICM can lead to cold gas ($\sim$1 keV) reservoirs around the BCG (e.g., \citealt{2001A&A...365L.104P,2004cgpc.symp..143D,2006PhR...427....1P}) - albeit lower than predicted from the standard cooling flow model of \citet{1994ARA&A..32..277F}. Imaging of the cluster cores revealed surface brightness depressions called X-ray cavities, which were found to be coincident with the lobes of the radio galaxy hosted in the BCG (see e.g., \citealt{1993MNRAS.264L..25B,2000A&A...356..788C,2000ApJ...534L.135M,2000MNRAS.318L..65F,2004ApJ...607..800B}). These works showed that the energy required to excavate the bubbles (in the range $10^{55 - 62}$ erg, e.g., \citealt{2014ApJ...786L..17V,2015ApJ...805...35H,2016ApJS..227...31S}) matches the energy released by cooling of the ICM; these evidences suggested that cluster central AGNs can prompt a deposition of energy in the ICM, preventing an over-cooling of the same gas that fuels their supermassive black holes (SMBH) (e.g., \citealt{2007ARA&A..45..117M,2012NJPh...14e5023M}). This finely-tuned balance is usually referred to as AGN feedback cycle. 
\\ \indent The end-product of ICM cooling is predicted to form reservoirs of multi-phase gas surrounding the BCG. Warm gas ($\sim10^{4}$ K) glowing in line emission (e.g., H$\alpha$, see e.g., \citealt{2010ApJ...721.1262M,2016MNRAS.460.1758H}) and molecular gas detected in CO ($\sim$10 - 100 K, see e.g., \citealt{2001MNRAS.328..762E,2018arXiv180707027C,2019MNRAS.490.3025R}) have been found to be co-spatial with the lowest entropy ICM. The multi-temperature medium is usually filamentary, and is thought to generate from condensation of the hot cluster gas (e.g., \citealt{2010ApJ...721.1262M,2020NatAs...4...10G}). Furthermore, several combined X-ray, radio and optical/millimeter studies have discovered that the cold filaments are draped around or trail X-ray cavities (e.g., \citealt{2012MNRAS.421.3409H,2015ApJ...811..111M,2019A&A...631A..22O,2019MNRAS.485..229R}). This coincidence has led to theories that the link between AGN activity and ICM cooling consists not only in lowering the amount of fuel, but also in setting the conditions for condensation to occur (e.g., \citealt{2015ApJ...802..118B,2016ApJ...829...90Y}). Besides heating the surrounding medium, the expansion and rise of X-ray cavities can drag enriched, cold material from the central regions up to several tens of kpc (e.g., \citealt{2011ApJ...732...13G}). By lifting low temperature gas to an altitude where cooling times and dynamical times become competitive, the condensation of the ICM into warm/cold gas kernels is stimulated (e.g., \citealt{2017ApJ...837..149G}).
\\ \indent In this context, it is interesting to note that in roughly 2/3 of cool core clusters there are surface brightness edges (named \say{cold fronts}) wrapped around the center in a spiral morphology \citep{2010A&A...516A..32G,2016JPlPh..82c5301Z}. Additionally, these system can show spatial offsets between the BCG and the coldest phase of the ICM. The mechanism generating these features is \textit{sloshing} of the gas in the potential well of the cluster, triggered by gravitational perturbations of the system (for reviews see \citealt{2007PhR...443....1M,2013ApJ...762...78Z,2016JPlPh..82c5301Z}); the oscillation of the cold gas leads to the formation of cold fronts on large scale, and can separate the peak of ICM emission from the AGN in the central regions. Recent studies have shown that sloshing can be responsible for offsetting not only the hot ICM (T$\sim$10$^{7-8}$ K), but also the multi-phase gas surrounding the BCG (e.g., \citealt{2018arXiv180707027C,2018ApJ...853..177P,2019ApJ...870...57V}), generating asymmetries in the distribution of the warm (T$\sim$10$^{4}$ K) and cold (T$\sim$10-100 K) phases. As this component is believed to sustain the activity of the central AGN by feeding the SMBH, a recent interest in the effect of sloshing on the feedback cycle has been growing. If sloshing separates the AGN and its fuel reservoirs, then the stability of feeding and feedback might be affected. Early results suggest that sloshing does not break the feedback cycle, but could possibly influence the timescales of AGN activity (as in e.g., Abell~1991, \citealt{2012MNRAS.421.3409H}; Abell~2495, \citealt{2019ApJ...885..111P}; Abell~1668, \citealt{pasini2021}).

\subsection{The galaxy cluster ZwCl~235}\label{subsec:z235}
\begin{table}[ht]
	\centering
	\renewcommand{\arraystretch}{1.3}
	\caption{Systems with X-ray flux greater than $9\times10^{-12}$ erg cm$^{-2}$ s$^{-1}$ from the sample of \citet{1998MNRAS.301..881E}, and with an H$\alpha$ luminosity greater than $10^{40}$ erg s$^{-1}$ from the sample of \citet{1999MNRAS.306..857C}, ordered by decreasing H$\alpha$ luminosity.}
	\label{tab:objects}
	\begin{tabular}{l|c|c|c}
		\hline
		 Object & $f_{X}$ & $L_{H\alpha}$ & Literature \\
		        & [erg cm$^{-2}$ s$^{-1}$] & [erg s$^{-1}$] & (e.g.)\\
		\hline
		 Abell~1068 &  9.4 $\times10^{-12}$ & 172.3 $\times10^{40}$ & (1), (2) \\
		 \hline
	   	 Abell~1835 & 14.7 $\times10^{-12}$ & 163.9 $\times10^{40}$ & (3), (4)\\
	   	 \hline
	   	 Abell~2204 & 21.9 $\times10^{-12}$ & 159.4 $\times10^{40}$ & (5), (6)\\
	   	 \hline
	   	 Abell~2390 &  9.6 $\times10^{-12}$ & 61.6 $\times10^{40}$ & (7), (8)\\
	   	 \hline
	   	 RXJ1720+26 & 14.3 $\times10^{-12}$ & 12.7 $\times10^{40}$ & (9) \\
	   	 \hline
	   	 Abell~115  &  9.0 $\times10^{-12}$ & 12.7 $\times10^{40}$ & (10) \\
	   	 \hline
	   	 ZwCl~8276  & 16.4 $\times10^{-12}$ & 12.5 $\times10^{40}$ & (11) \\
	   	 \hline
	   	 Abell~1795 & 68.1 $\times10^{-12}$ & 11.3 $\times10^{40}$ & (12), (13)\\
	   	 \hline
	   	 Abell~478  & 39.9 $\times10^{-12}$ & 10.8 $\times10^{40}$ & (14)\\
	   	 \hline
	   	 2A0335+096 & 80.5 $\times10^{-12}$ & 10.3 $\times10^{40}$ & (15), (16)\\
	   	 \hline
	   	 Abell~2009 &  9.2 $\times10^{-12}$ & 6.1 $\times10^{40}$ & (...) \\
	   	 \hline
	   	 \textcolor{red}{ZwCl~235}   & 10.9 $\times10^{-12}$ & 4.1 $\times10^{40}$ & (...) \\
	   	 \hline
		 Abell~2199 & 96.8 $\times10^{-12}$ & 2.7 $\times10^{40}$ & (17)\\
		 \hline
		 Abell~2052 & 47.1 $\times10^{-12}$ & 2.6 $\times10^{40}$ & (18), (19)\\
		 \hline
		 Abell~1668 &  9.3 $\times10^{-12}$ & 2.3 $\times10^{40}$ & (20) \\
		 \hline
		 Abell~2495 & 11.8 $\times10^{-12}$ & 2.0 $\times10^{40}$ & (21)\\
		 \hline
		 Abell~2634 & 23.1 $\times10^{-12}$ & 1.3 $\times10^{40}$ & (22), (23)\\
		 \hline
		 Abell~1991 &  9.4 $\times10^{-12}$ & 1.1 $\times10^{40}$ & (24), (25)\\
        \hline
	\end{tabular}
	\tablefoot{The target of our study (ZwCl~235) is highlighted in red. (1) Object name; (2) X-ray flux; (3) H$\alpha$ luminosity; (4) Examples of relevant literature studies dedicated to the object: (1) \citet{2004ApJ...601..173M}; (2) \citet{2004ApJ...601..184W}; (3) \citet{mcnamara2006}; (4) \citet{mcnamara2014}; (5) \citet{2009MNRAS.393...71S}; (6) \citet{2017ApJ...838...38C}; (7) \cite{2001MNRAS.324..877A}; (8) \citet{sonkamble2015}; (9) \citet{2001ApJ...555..205M}; (10) \citet{2018ApJ...859...44H}; (11) \citet{ettori2013}; (12) \citet{kokotanekov2018}; (13) \citet{russell2017}; (14) \citet{2003ApJ...587..619S}; (15) \citet{2009MNRAS.396.1449S}; (16) \citet{2016ApJ...832..148V}; (17) \citet{2013ApJ...775..117N}; (18) \citet{blanton2011}; (19) \citet{balmaverde2018}; (20) \citet{pasini2021}; (21) \citet{2019ApJ...885..111P}; (22) \citet{schindler1997}; (23) \citet{2007ApJ...657..197S}; (24) \citet{sharma2004}; (25) \citet{hamer2012}.}
\end{table}
To explore which conditions trigger the formation of a multi-phase medium around BCGs, and how the dynamics of the environment can influence AGN feeding and feedback, systems with a relatively high X-ray flux and bright in H$\alpha$ (suggesting the presence of cool gas) can be considered.  
To identify a potentially interesting system, we selected objects with X-ray flux greater than $9.0\times10^{-12}$ erg cm$^{-2}$ s$^{-1}$ from the BCS survey of \citet{1998MNRAS.301..881E}, and with an H$\alpha$ luminosity greater than $10^{40}$ erg s$^{-1}$ from the sample of \citet{1999MNRAS.306..857C} (for a similar selection see \citealt{ettori2013,2019ApJ...885..111P,pasini2021}).
The 18 objects satisfying these criteria are listed in Tab. \ref{tab:objects}. Many clusters in the list (e.g., Abell~2052, 2A0335+096, Abell~1835, Abell~1795, Abell~1991 and Abell~2199) provide archetypal examples of the interplay between the central AGN and the surrounding gas for the detection of X-ray cavities, for the existence of filamentary warm nebulae, and/or for the effect of sloshing on the central cool gas. Other objects, such as Abell~2495 and Abell~1668 (characterized by a relatively low H$\alpha$ luminosity, see Tab. \ref{tab:objects}), have been only recently investigated: the analysis of the two revealed evidence for a sloshing-influenced AGN feedback cycle - with these clusters having the X-ray peak, the BCG and the warm H$\alpha$-emitting gas phase offset from each other \citep{2019ApJ...885..111P,pasini2021}. As an informative remark, we note that a similar X-ray flux selection in e.g., the REFLEX sample \citep{2004A&A...425..367B} identifies other objects (e.g., S1101, Hydra~A, Abell~85, Abell~133, Abell~496, Abell~2597, NGC~5044, Centaurus) that represent further notable laboratories where the AGN-ICM interplay has been extensively investigated.
\\ \indent In this work we progress on the study of X-ray and H$\alpha$ bright systems by focusing on ZwCl~235, one of the two clusters in Tab. \ref{tab:objects} that still lacks a dedicated study (the other being Abell~2009). ZwCl~235 is a nearby system ($z\sim0.083$) located at RA, DEC = 00:43:52.0, +24:24:21 (J2000), with $f_{X} = 1.1\times10^{-11}$ erg cm$^{-2}$ s$^{-1}$, $L_{X}^{0.1-2.4\,\text{keV}}=3.22\times10^{44}$ erg s$^{-1}$, and $L_{H\alpha}=(4.1\pm0.6)\times10^{40}$ erg s$^{-1}$. The cluster has a 20 ks observation in the \textit{Chandra} data archive \footnote{\url{https://cxc.harvard.edu/cda/}} that allows us to investigate the ICM properties. However, the high spatial resolution multi-wavelength coverage of ZwCl~235 is scarce (compared to the well-studied objects at the top of Tab. \ref{tab:objects}). This source is twelfth in H$\alpha$ luminosity in the list of 18 sources, indicating a relatively low star formation rate. This is consistent with the sub-millimeter IRAM observations presented in \citet{2003A&A...412..657S}, that found hints of a CO(1-0) line emission in the core of ZwCl~235 and failed to detect CO(2-1), thus placing an upper limit on the molecular mass of M$_{\text{mol}}\leq2.5\times10^{9}$ M$_{\odot}$ (in the same work, molecular masses of more than $4\times10^{10}$ M$_{\odot}$ are measured in Abell~1068 and Abell~1795). The relatively low level of star formation is also evident from the upper limit on the infrared luminosity set by \citet{2008ApJS..176...39Q}, a feature shared with  Abell~2009, Abell~1668, Abell~2495 and Abell~1991, i.e. the systems in Tab. \ref{tab:objects} with an H$\alpha$ luminosity similar to or below that of ZwCl~235. The radio emission of the BCG in ZwCl~235 between 1 GHz - 150 GHz was analyzed in \citet{2015MNRAS.453.1223H,2015MNRAS.453.1201H}, who found that a single power-law of flat spectral index $\alpha=-0.45$ provides a good description of the radio spectral energy distribution, indicating a core-dominated source. Faint and extended - but barely resolved - radio lobes connected to the bright core are visible in the survey LOFAR image at 144 MHz shown in \citet{2020MNRAS.496.2613B}. Since dedicated, spatially-resolved radio observations of the AGN in ZwCl~235 are missing, we employ survey radio images to further examine the radio morphology of the BCG. Overall, our aim is to investigate the X-ray/radio properties of ZwCl~235 and discuss the results in light of the optical/sub-millimeter information mentioned above, in order to draw a picture of AGN feeding and feedback in this system.
\\ \indent  This paper is organized as follows: Section \ref{datamethod} describes the radio and X-ray data used in this work, summarizing the main steps of data reduction. The analysis and results and reported in Section \ref{result}, with dedicated subsections to present in particular: the radio properties of the BCG (\cref{radiogalaxy}), the general X-ray analysis of the cluster (\cref{general}, \cref{fronts}), the radio and X-ray evidence for AGN feedback (\cref{feedback}), the peculiarities of the metal distribution in the ICM (\cref{metal}), the X-ray emission from the BCG (\cref{bcg}), and the properties of the ICM in the central 15 kpc (\cref{brightf}). We discuss our results in Section \ref{discussione}, and summarize our main conclusions in Section \ref{conclusione}.
\\ \indent We adopt the following cosmology: $H_{0}$ = 70 km s$^{-1}$ Mpc$^{-1}$, $\Omega_{\text{m}}$ = 0.3, $\Omega_{\Lambda}$ = 0.7, which results in a conversion between linear and angular scales of 1.56 kpc/'' at the redshift of ZwCl~235 (corresponding to a luminosity distance of $D_{L}=378$ Mpc). We report every uncertainty at the $1\sigma$ confidence level. The radio spectral index $\alpha$ is defined as $S_{\nu}\propto\nu^{\alpha}$ (where $\nu$ is the frequency and $S_{\nu}$ is the flux density).
\section{Data and Methods}
\label{datamethod}
\subsection{Radio data - \textit{LOFAR, VLA, VLBA}}
\begin{figure}[ht]
\includegraphics[width=1\linewidth]{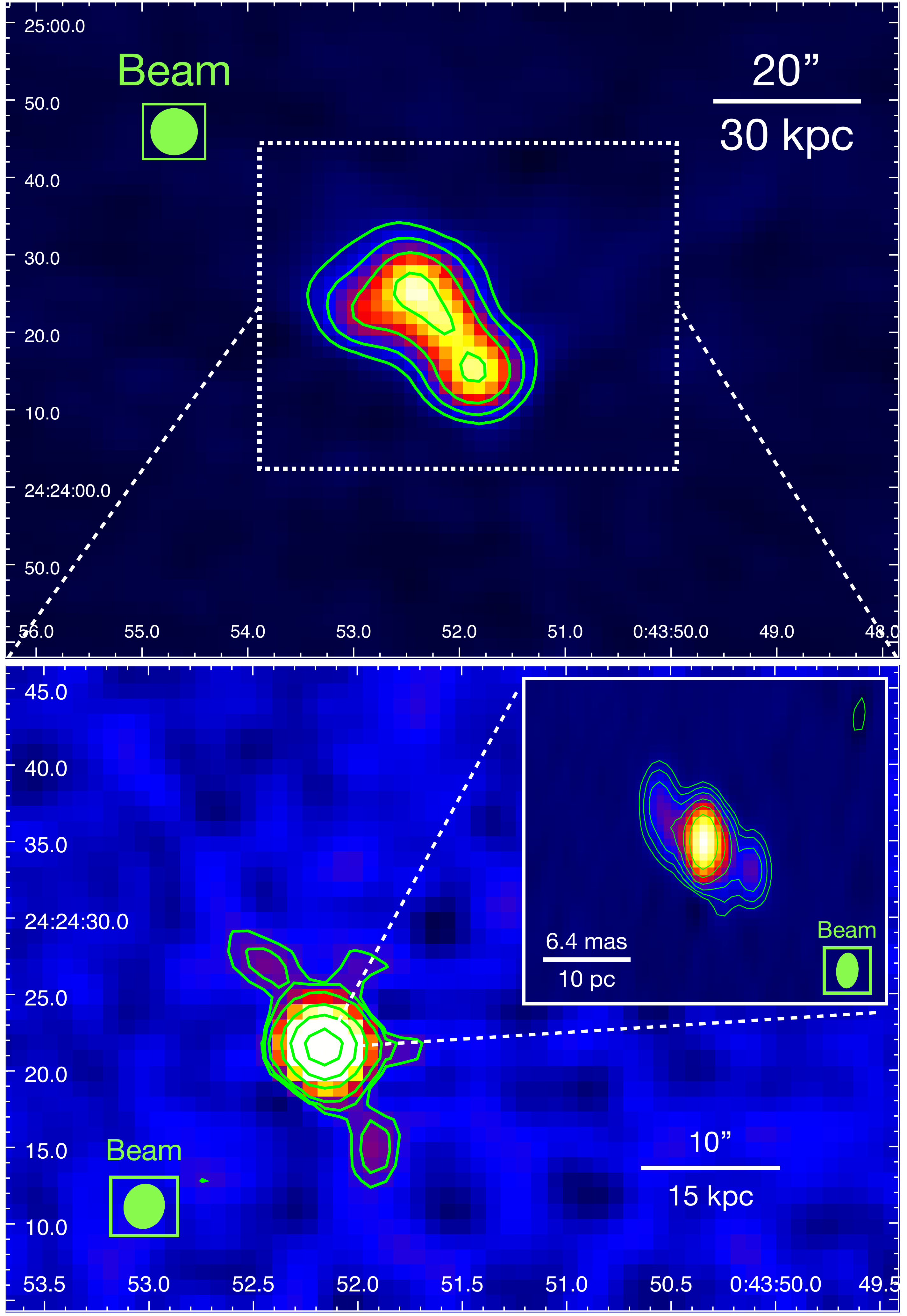}
\caption{Resolved radio observations of ZwCl~235. \textit{Upper panel:} 144 MHz LOTSS DR2 image of ZwCl~235, at a resolution of 6$''$. Green contours are drawn at 5, 10, 20, 40 $\times\sigma_{rms}$ (with $\sigma_{rms}=0.1$ mJy beam$^{-1}$). \textit{Lower panel:} 3 GHz VLASS image of the radio galaxy in the BCG, at a resolution of 2.7$''\times2.5''$. Green contours are drawn at 3, 6, 12, 36, 72, 144, 288 $\times\sigma_{rms}$ (with $\sigma_{rms}=0.2$ mJy beam$^{-1}$). \textit{Lower zoom-in:} 5 GHz VLBA image of the core, at a resolution of 3.2$\times1.3$ mas. Green contours are drawn at 4, 8, 16, 32, 64, 128$\times\sigma_{rms}$ (with $\sigma_{rms}=26\,\mu$Jy beam$^{-1}$). For each panel, the beam is represented by a green circle.
              }
         \label{figcav_paper}
\end{figure}

In order to derive the properties of the central radio galaxy and perform comparisons with the X-ray observation, we employed radio observations at different frequencies and with different resolutions.
\\ \textit{LOFAR - 144 MHz.} The recent release of the LOFAR Two-meter sky survey DR2 \citep{2022arXiv220211733S} allows us to access the LOFAR observation of ZwCl~235, originally published in \citet{2020MNRAS.496.2613B}. The 144 MHz total intensity image has been retrieved from the archive \footnote{\url{https://lofar-surveys.org/dr2_release.html}} and is shown in Fig. \ref{figcav_paper}. The resolution is 6$''\times6''$, and the r.m.s. noise is $\sigma_{rms}=$0.1 mJy beam$^{-1}$. 
\\ \textit{VLA - 3 GHz.} ZwCl~235 has also been observed as part of the VLA Sky Survey at 3 GHz. The observation is part of the second campaign of the first survey epoch (VLASS1.2, \citealt{2020PASP..132c5001L}), and the target falls in the tile T17t01. The 3 GHz total intensity image has been retrieved from the archive \footnote{\url{science.nrao.edu/vlass/data-access}} and is shown in Fig. \ref{figcav_paper}. The resolution is 2.7$''\times2.5''$ (position angle PA=162$^{\circ}$), and the r.m.s. noise is $\sigma_{rms}=$0.2 mJy beam$^{-1}$. The fluxes (within 3$\sigma_{rms}$) of the different components have been measured in CASA 6.1.2 (see \cref{radiogalaxy}).
\\ \textit{VLBA - 5 GHz.} The source has been targeted with the VLBA at 5 GHz in phase-referencing mode for 45 minutes. \citet{2015MNRAS.453.1201H} reported that at milli-arcsecond (mas) scale the source is extended, showing a two-sided jet structure. In order to perform morphological comparisons with the VLA data, we obtained the VLBA data at 5 GHz from the archive \footnote{\url{https://data.nrao.edu}}. Standard data reduction was performed in AIPS v.31DEC20. The VLBA data were imaged in AIPS using the task \texttt{IMAGR} and setting the robust parameter to 1, obtaining total and peak flux measurements consistent with those of \citet{2015MNRAS.453.1201H}. The resulting radio map, at a resolution of 3.2$\times1.3$ mas (r.m.s noise $\sigma_{rms}=26\,\mu$Jy beam$^{-1}$), is shown in Fig. \ref{figcav_paper}. 

\subsection{X-ray data - \textit{Chandra}}
ZwCl~235 was observed by \textit{Chandra} using ACIS-S during Cycle 11 (ObsID 11735), for a total, uncleaned exposure of 19.8 ks. We retrieved the data from the \textit{Chandra} archive\footnote{\url{cda.harvard.edu/chaser}} and reprocessed the observation using \texttt{CIAO-4.13}. Point sources were identified using the tool \texttt{wavdetect}, and masked during the morphological and spectral analysis. By cross-matching the detected X-ray point sources with the USNO-A.2 optical catalog we verified that the astrometry of the \textit{Chandra} observation does not need further corrections beyond its nominal pointing accuracy (0.4$''$). We filtered the data from periods contaminated by background flares, obtaining a cleaned exposure of 19.5 ks. The blank-sky event file matching the observation of ZwCl~235 has been selected as background file, and normalized by the 9-12 keV count-rate of the observation. 
\\ \indent In order to investigate in details the morphology of the ICM, we produced images and corresponding background and exposure maps in the 0.5 - 7 keV energy band. These images have been used to extract surface brightness profiles, which were fit using the software \texttt{Proffit-1.5} \citep{2011A&A...526A..79E}.
\\ \indent Spectral fitting has been performed in the 0.5 - 7 keV band using \texttt{XSPEC v.12.10}, binning the spectra to 25 counts per bin and applying the $\chi$-statistics (unless otherwise stated). For every model discussed in this article, we employed the table of solar abundances of \citet{2009ARA&A..47..481A}. When fitting the spectra, we always included an absorption model (\texttt{tbabs}) to account for Galactic absorption, with the column density fixed at the value $N_{\text{H}} = 3.57 \times 10^{20}$ cm$^{-2}$ \citep{2016A&A...594A.116H}. When included in the fitting model, the redshift $z=0.083$ was also fixed.

\section{Results}
\label{result}
In the following we describe the results of the morphological and spectral analysis of the radio and X-ray data of ZwCl~235 and of the central AGN.

\subsection{The central radio galaxy}
\label{radiogalaxy}
In this section we discuss the morphology and spectral properties of the radio galaxy associated with the BCG of ZwCl~235, combining our new results with those presented in the literature.
\\ \indent  In the \textit{upper panel} of Fig. \ref{figcav_paper} we show the 144 MHz LOFAR observation of the cluster, that reveals extended radio lobes (each $\sim$20 kpc long) oriented north-east (NE) to south-west (SW), with a position angle of $\sim$130$^{\circ}$. Using this observation, \citet{2020MNRAS.496.2613B} measured the source total flux density $S_{\nu}=151\pm24$ mJy. 
\\ \indent  The lower panel of Fig. \ref{figcav_paper} shows the VLASS 3 GHz image of the radio galaxy at the center of ZwCl~235. The morphology is slightly resolved, showing a bright core, two opposite radio lobes extending NE to SW for roughly 15 kpc, and two small west protrusions. The source has a total flux density of 37.7$\pm$1.0 mJy, with a contribution of 32$\pm$0.9 mJy coming from the bright core (which supports the result of \citet{2015MNRAS.453.1201H} that the total flux of this radio galaxy is dominated by the core emission). We note the similar morphology between the 3 GHz image and the 144 MHz one, as both unveil the presence of radio lobes with a position angle of $\sim$130$^{\circ}$. 
\\ \indent Exploring radio emission on the parsec scale can provide further insights on the AGN activity of a BCG (e.g., \citealt{2010A&A...516A...1L}). To this end, in the zoom-in of Fig. \ref{figcav_paper} we show the VLBA 5 GHz image of the radio galaxy's core at a resolution of 3.1$\times$1.4 mas$^{2}$ ($\sim$4.7$\times$2.1 parsec$^{2}$). We recover an unresolved core plus a two sided-jet structure oriented NE to SW (consistent with \citealt{2015MNRAS.453.1201H}). Each jet has a projected length of $\approx$10 pc. The position angles of the jets, measured with a straight line passing through the core that connects the two jets, is $\sim$140$^{\circ}$, which is in good agreement with that of the radio galaxy on kpc-scale. Therefore, the VLBA image likely unveils the details of the jets that have inflated the radio lobes seen at 3 GHz and 144 MHz. Moreover, the bright core visible in the VLBA data indicates that the central engine of the BCG is currently active (consistent with the aforementioned core dominance). By placing the core radio properties of ZwCl~235 in the context of other BCGs, we can specifically determine how active the source is with respect to the general population. In particular, we refer to the results of \citet{2015MNRAS.453.1201H}, who determined from the radio spectrum that the core component of this radio galaxy is expected to have a radio power of roughly $4\times10^{23}$ W/Hz at 10 GHz, and put an upper limit to the expected radio power at 1 GHz of any extended emission of $\leq1.2\times10^{24}$ W/Hz. On the one hand, the 10 GHz core radio power places ZwCl~235 in the top 20\% of the sample of \citet{2015MNRAS.453.1201H}, which indicates that the central AGN is currently more active than most other BCGs. On the other hand, the upper limit on the 1 GHz radio power of extended components is close to the average of the sample, suggesting that on longer timescales the activity of the central radio galaxy is typical of BCGs. In this respect, restricting the comparison to the X-ray flux-limited list of Tab. \ref{tab:objects}, ZwCl~235 is similar to ZwCl~8276, another core dominated source (see \citealt{2015MNRAS.453.1201H,2015MNRAS.453.1223H}).
\\ \indent  The availability of the LOFAR and VLASS data, that reveal extended lobes of the radio galaxy previously unknown, enables us to constrain the spectral properties of the extended emission by considering the relative contribution of the core flux to the total flux. Using an archival, snapshot VLA observation at 4.8 GHz of ZwCl~235 (which did not resolve the source, having a beam of $\sim5''$ FWHM), \citet{2015MNRAS.453.1201H} measured a total flux of 30.3 mJy and a core peak flux of 27.3 mJy (88\% of which is recovered by the VLBA observation, see \citealt{2015MNRAS.453.1201H}). Combining these information at 4.8 GHz with our measurements at 3 GHz and the total flux measured by LOFAR at 144 MHz, we estimate the spectral index of the core and extended components of the radio galaxy with the following methods:
\begin{enumerate}
    \item \textit{Unresolved core}. The $32$ mJy at 3 GHz from the VLASS data and 27.3 mJy at 4.8 GHz from the VLA data suggest a core spectral index of $\alpha^{4.8}_{3}\approx-0.3$. This rather flat value is in good agreement with the expected emission from a core: in \citet{2015MNRAS.453.1201H} the average core spectral index of BCGs is $-0.33$; limiting ourselves to the sources listed in Tab. \ref{tab:objects} we find $\overline{\alpha}=-0.31$.
    \item \textit{Extended components at GHz frequencies}. The subtraction of the core flux from the total flux can provide an estimate of the amount of emission coming from extended components. With residual fluxes of $5.7$ mJy at 3 GHz and 3 mJy at 4.8 GHz, the extended component spectral index measured from the VLA and VLASS data is $\alpha^{4.8}_{3}\approx-1.3$. A steep spectral index ($\alpha\leq-1$) is usually associated with the ageing of the electron population responsible for the radio emission.
    \item \textit{Extended components between MHz and GHz frequencies.} Following the procedure described in \citet{2021MNRAS.503.4627U}, we estimated the relative contribution of the core and the extended components to the total flux at 144 MHz ($\approx$151 mJy, \citealt{2020MNRAS.496.2613B}). Assuming that the core spectral index remains flat at lower frequencies, we extrapolated the AGN flux at 3 GHz to 144 MHz, finding $S^{144\,MHz}_{\text{core}}\approx73$ mJy and thus $S^{144\,MHz}_{\text{extended}} = S^{144\,MHz}_{\text{total}} - S^{144\,MHz}_{\text{core}}\approx78$ mJy. This corresponds to a 144 MHz - 3 GHz spectral index of the extended components of $\alpha^{3}_{0.144}\approx-0.87$.
\end{enumerate}
Overall, our estimates support a steepening beyond 3 GHz of the spectral index of the extended components detected by the VLASS and the VLA. Instead, regarding the core emission, \citet{2015MNRAS.453.1223H} reported a possible steepening of the spectrum only beyond 100 GHz (with fluxes at 16 GHz, 90 GHz and 150 GHz of 16.7 mJy, 9.9 mJy and 4.7 mJy, respectively). We note that our estimates above for the extended components should be treated with caution, as a proper measurement of spectral index would require observations with similar \textit{u-v} coverage, or at least to resolve the same structures. If confirmed by future, more sensitive and resolved radio observations, the above results would be consistent with ageing of the radio lobes. Using the available data and adopting a few assumptions, it is possible to derive only a tentative upper limit on the radiative age of the lobes. In particular, given the spectral break frequency $\nu_{b}$ [GHz] and the magnetic field of the radio galaxy $B$ [$\mu$G], it is possible to obtain the radiative age of a synchrotron emitting source (e.g., \citealt{2014NJPh...16d5001E}):
\begin{equation}
\label{tsync}
    \tau_{syn}\text{[Myr]} = 1590 \frac{B^{1/2}}{B^{2}+B^{2}_{IC}} [\nu_{br}(1+z)]^{-1/2}
\end{equation}
\noindent where $B_{IC} = 3.2(1+z)^{2}\,\mu$G is the Cosmic microwave background equivalent magnetic field. Due to the steepening of the spectral index of extended emission on the VLA-LOFAR kpc scales beyond 3 GHz, we can assume $\nu_{br}=3$ GHz. Since the flux of extended components from the snapshot VLA observation at 4.8 GHz is likely to be a lower limit, it is possible that the break frequency could be higher than 3 GHz. Therefore, the derived $\tau_{syn}$ has to be treated as an upper limit. In light of this consideration, we used the minimum energy loss field $B=B_{IC}/\sqrt{3}=2.17\,\mu$G (see \citealt{2017SciA....3E1634D}), that minimizes the radiative losses of the emitting particles. This second assumption provides another upper limit on the radiative lifetime, thus it is consistent with selecting $\nu_{br}=3$ GHz. By substituting these values in Eq. \ref{tsync} we obtained $\tau_{syn}\lessapprox70$ Myr.

\subsection{Global X-ray properties of the ICM: radial profiles and spectral maps}
\label{general}
\begin{figure*}
\includegraphics[width=1\linewidth]{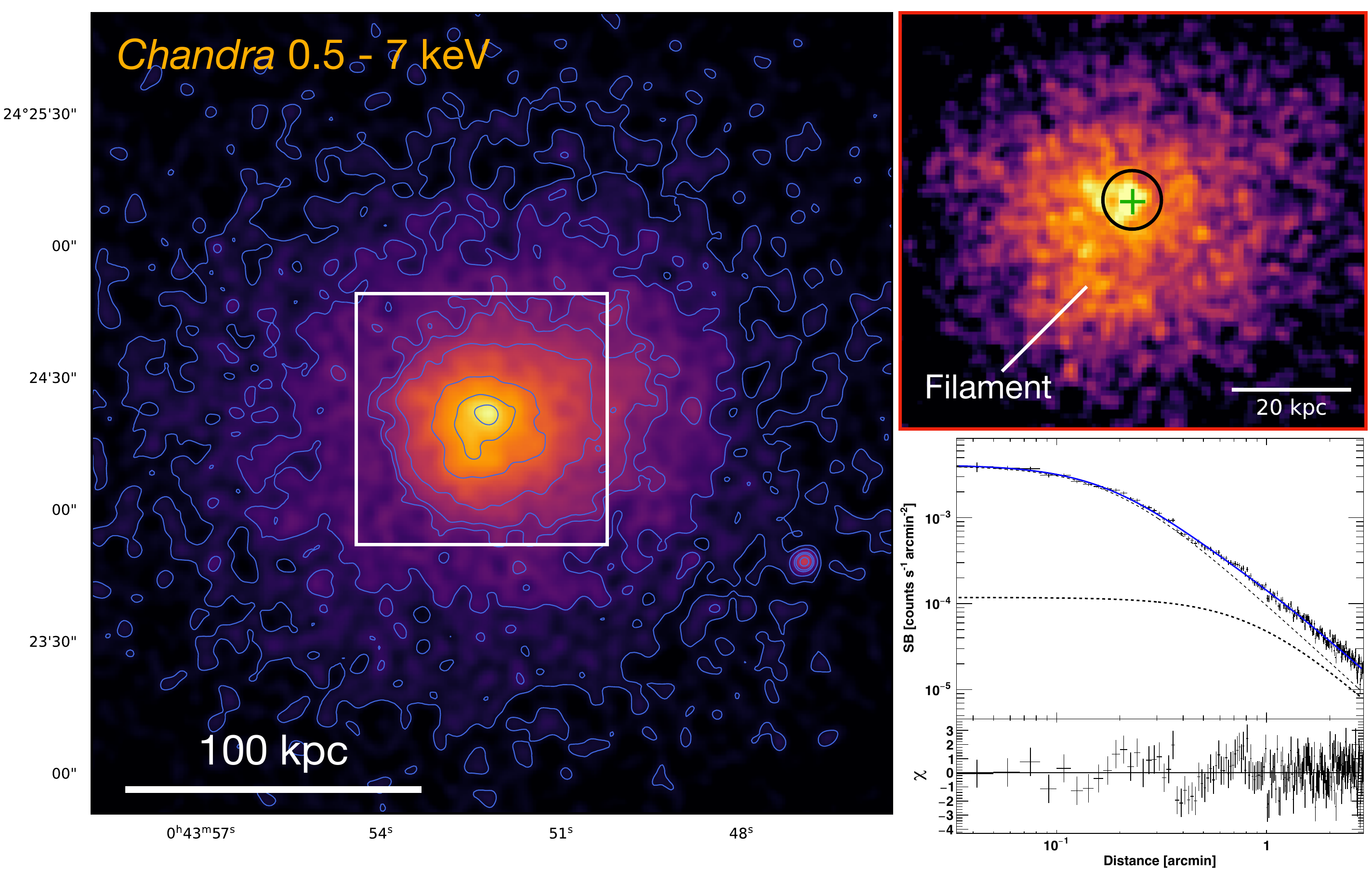}
\caption{\textit{Left panel:} 0.5 -7 keV band background subtracted, exposure corrected \textit{Chandra} image of ZwCl~235, centered on the BCG. The image is smoothed with a Gaussian of kernel size 3$''$. Blue contours are spaced by a factor of 1.5, with the highest being 6$\times10^{-7}$ cts s$^{-1}$ cm$^{-2}$. The white box indicates the size of the zoom-in on the right. \textit{Right upper panel:} Zoom-in of the \textit{Chandra} image in the left panel, smoothed with a Gaussian of kernel size 1.5$''$. The green cross marks the BCG center, and the black circle shows the size of the region excised from the surface brightness radial profile. The position of the bright filament discussed in \cref{general} and \cref{brightf} is indicated. \textit{Right lower panel:} Surface brightness profile of ZwCl~235, with the best fit double $\beta$-model overplot with a blue line (dotted black lines represent the contributions of the two $\beta$-models). The residuals are shown in the lower box.}
         \label{figchandra}
\end{figure*}
\begin{figure*}[ht]
\centering
\includegraphics[width=0.9\linewidth]{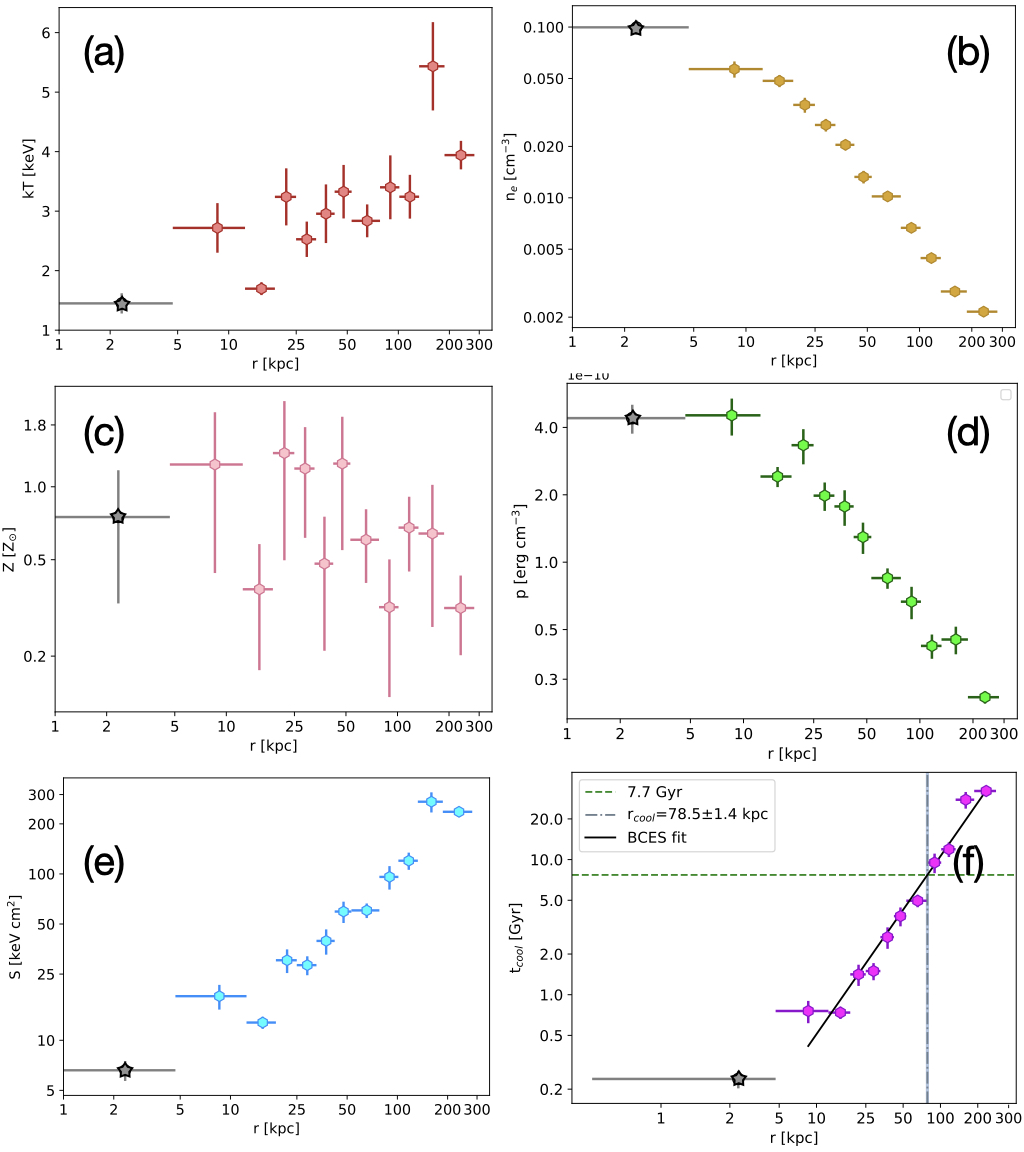}
\caption{ Circles: deprojected thermodynamic profiles of the ICM in ZwCl~235 between 3$''$ - 180$''$ (for details on the spectral extraction and fitting see \cref{general}); Star: the innermost gray point in each plot is obtained from the spectral analysis of the inner 3$''$ of the cluster (for details see \cref{bcg}). \textit{Panel a:} temperature profile. \textit{Panel b:} electron density profile. \textit{Panel c:} abundances profile. \textit{Panel d:} pressure profile. \textit{Panel e:} entropy profile. \textit{Panel f:} cooling time profile, with the best fit power-law over the radial range 3$''$ - 180$''$ overlaid in black, the t$_{\text{cool}}$=7.7 Gyr threshold shown with a green dashed line and the cooling radius shown with a grey line. The fit has been performed using BCES \citep{1996ApJ...470..706A}.
              }
         \label{profile}
\end{figure*}
\begin{figure*}
\includegraphics[width=1\linewidth]{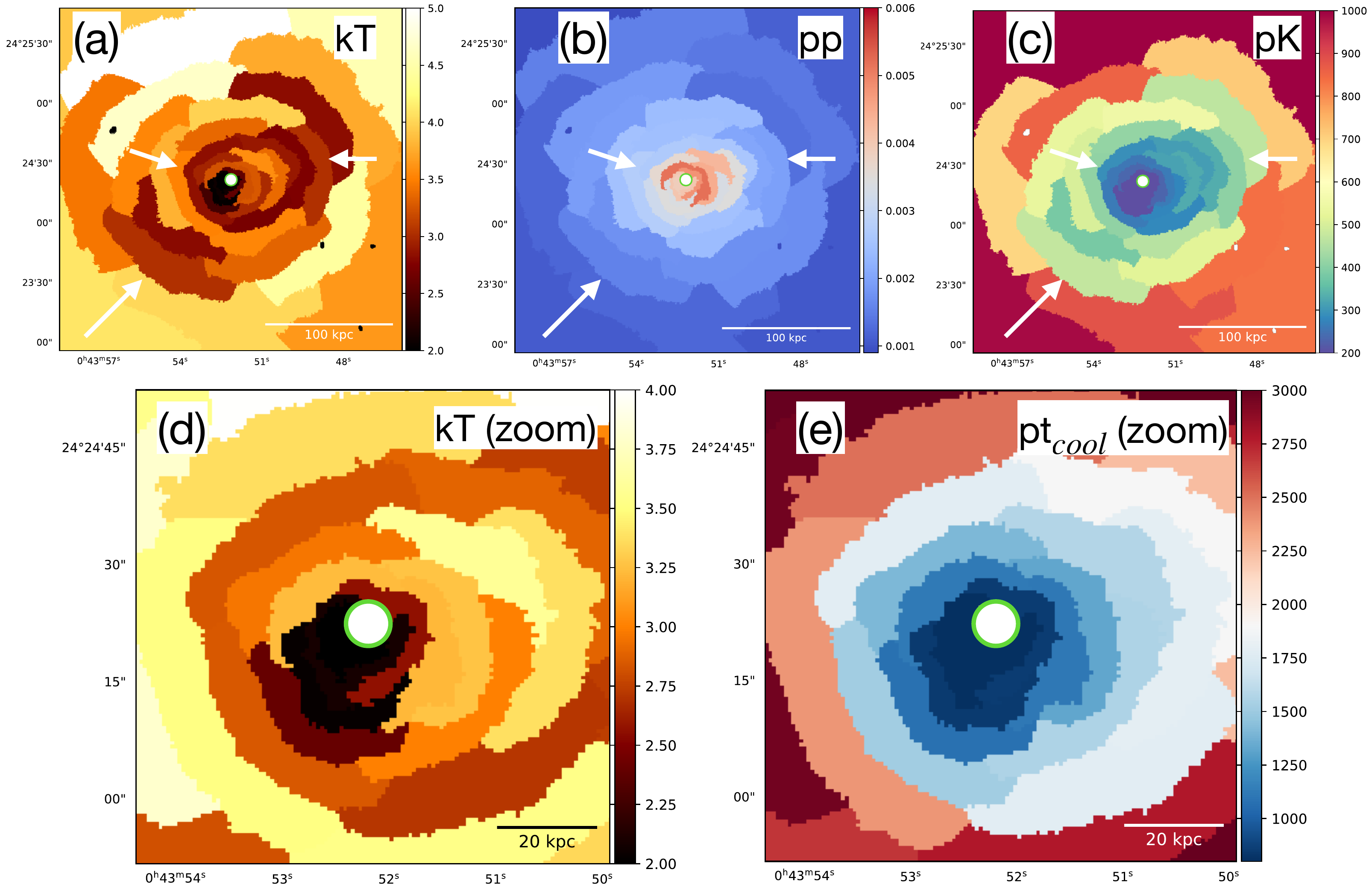}
\caption{Spectral maps of the ICM in ZwCl~235 (see Sect \ref{general} for details). In the \textit{upper panels}, the temperature (keV), \textit{pseudo}-pressure (arbitrary units) and \textit{pseudo}-entropy (arbitrary units) maps are shown. Relative errors are of the order of $\sim$20\%. The overlaid white arrows mark the positions of the surface brightness edges discussed in \cref{fronts}. In the \textit{lower panels}, the zoomed temperature map (keV) and the \textit{pseudo}-cooling time map (arbitrary units) show the distribution of the cold gas in the inner $\sim$50 kpc. In all panels, the white-filled, green circle marks the extent of the central X-ray source (see \cref{bcg}).
              }
         \label{figspectral}
\end{figure*}

We show in Fig. \ref{figchandra} (\textit{left panel}) the 0.5 - 7 keV \textit{Chandra} image of ZwCl~235. The cluster is fairly spherical, with a bright edge to the east of the core. The zoom-in of Fig. \ref{figchandra} (\textit{right upper panel}) shows a detail of the cluster core, where it is possible to see the X-ray peak coinciding with the center of the BCG (green cross). We note the presence of a bright arm-like feature that, starting from the X-ray peak, bends south-east and reaches a distance from the center of 15$''$ ($\sim$23 kpc). \\ \indent  We extracted an azimuthally-averaged surface brightness profile from circular annuli with 1$''$ ($\sim$1.6 kpc) of bin width, extending from 3$''$ (to avoid any contamination from possible non-thermal X-ray emission related to the central AGN - see \cref{bcg} for the analysis of the inner 3$''$) to 180$''$ ($\sim$280 kpc) from the center (shown in Fig. \ref{figchandra}, \textit{right lower panel}). The profile is typical of cool core galaxy clusters, as we observe no internal flattening of the profile, but rather a continuous increase in surface brightness towards the center. The profile has been fit with a single $\beta$-model and a double $\beta$-model (with a common $\beta$ value). We found that the double $\beta$-model with a $\chi/$D.o.f = 195/166 provides a better description of the surface brightness profile w.r.t the single $\beta$-model, with a $\chi/$D.o.f = 205/168 (F statistics value of 4.3, \textit{p}-value of 0.016). This is typical of cool core galaxy clusters displaying an excess surface brightness w.r.t. the single $\beta$-model (e.g., \citealt{1999ApJ...517..627M}). The best fit parameters are $\beta$ = 0.54$\pm$0.02, inner core radius $r_{1}$ = 11.4$''\pm$0.7$''$ (17.8$\pm$1.1 kpc), outer core radius $r_{2}$ = 53.4$''\pm$9.0$''$ (84.9$\pm$14.0 kpc), normalization $S_{1}$ = (4.02$\pm$0.13) $\times$ 10$^{-3}$ counts s$^{-1}$ arcmin$^{-2}$, and a ratio $S_{2}/S_{1}$ = 0.03.
\\ \indent  In order to study the thermodynamic properties of the cluster, we extracted the spectra of circular, concentric annuli centered on the BCG, starting from 3$''$ from the center to 180$''$ from the BCG. These annuli were chosen to contain at least 2000 counts, or a signal to noise ratio (SNR) of 40. The spectra were fit with a \texttt{projct$\ast$tbabs$\ast$apec} model, leaving the temperature, the metallicity and the normalization free to vary. A deprojected spectral analysis allows to obtain 3D estimates of thermodynamic quantities. In particular, assuming $n_{e}=1.2n_{p}$, the electron density of the ICM can be derived from the normalization (\textit{norm}) of the deprojected \texttt{apec} component using the following expression (e.g., \citealt{2012AdAst2012E...6G}): 
\begin{equation}
\label{normne}
    n_{\text{e}} = \sqrt{10^{14}\bigg(\frac{4\pi \times norm \times[D_{\text{A}}(1+z)]^{2}}{0.83 V}\bigg)}
\end{equation}
\noindent where $D_{\text{A}}$ is the angular distance from the source and $V$ is the projected volume of the emitting region.
\\ \indent From the electron density and temperature of the ICM, we obtained the pressure $p$, the entropy $K$, and the cooling time $t_{\text{cool}}$ of the ICM, respectively defined as:
\begin{align}
\label{press}
    &p=1.83\,n_{e}\,kT \\
\label{entro}
    &K =\frac{kT}{n_{e}^{2/3}} \\
\label{coolt}
    &t_{\text{cool}} = \frac{\gamma}{\gamma -1} \frac{kT}{\mu \,X \,n_{\text{e}}\,\Lambda(T)}
\end{align}
\noindent where $\gamma = 5/3$ is the adiabatic index, $\mu\approx0.6$ is the molecular weight, $X\approx0.7$ is the hydrogen mass fraction and $\Lambda(T)$ is the cooling function \citep{1993ApJS...88..253S}. Fig. \ref{profile} shows deprojected temperature, density, abundance, pressure, entropy and cooling time profiles.
\\ \indent  Additionally, we generated spectral maps by binning the 0.5 - 7 keV image with the \texttt{CONTOUR BINNING} technique \citep{2006MNRAS.371..829S}, in order to obtain a list of regions with a SNR$\geq$30. Then, we extracted and fitted the spectrum of each region with a thermal model (\texttt{tbabs$\ast$apec}), and produced maps of best fit values for temperature, metallicity and normalization. From the temperature and normalization maps it is possible to obtain \textit{pseudo}-pressure ($pp$), \textit{pseudo}-entropy ($pK$) and \textit{pseudo}-cooling time maps ($pt_{\text{cool}}$). In particular, the emission measure ($EM$) of the plasma is proportional to the ratio between the \texttt{apec} normalization and the number of pixels in each spectral region (\textit{norm}/n$_{pix}$). Considering that the electron density is proportional to the root of the projected emission measure, the above \textit{pseudo}-maps can be built using the following equations, respectively: 
\begin{align}
& pp = kT\,\times EM^{1/2} \\
& pK = kT\,\times EM^{-1/3} \\
& pt_{\text{cool}} = kT^{1/2}\,\times EM^{1/2}
\end{align}
We show the $kT$, $pp$, $pK$ and $pt_{\text{cool}}$ maps in Fig. \ref{figspectral}. 
\\ \indent As a note of caution, we observe that in order to securely detect spectral features (especially metallicity) from thermodynamic maps, a SNR$\geq$50 is typically required (e.g., \citealt{2019MNRAS.488.2925O}). However, given the short exposure of the \textit{Chandra} observation of ZwCl~235, we reduced the minimum SNR to 30, in order to preserve good spatial resolution. This resulted in relative uncertainties that can reach $\sim$20\% on temperature and $\sim$30\% on metallicity. Therefore, every feature we tentatively identified from the maps has been subsequently studied in details by extracting and fitting the spectrum of larger regions encompassing the specific feature, thus reaching a higher SNR and more robust results. We note that both the profiles and spectral maps are typical of cool core galaxy clusters, showing a central increase in gas density and a decline in temperature. 
\\ \indent It has been argued that the use of \texttt{projct} to deproject spectra may generate unphysical oscillating temperature profiles \citep{2008MNRAS.390.1207R}. As a sanity check, we tested the alternative use of the \texttt{dsdeproj} code by \citet{2016ascl.soft10003S} to deproject the spectra, which returned consistent results with those of \texttt{projct}. 
\\ \indent  The standard method to determine the magnitude of cooling in a cool core galaxy cluster consists in measuring the bolometric X-ray luminosity associated with the ICM within the so-called \say{cooling radius}. It is possible to define the cooling radius as the region within which the cooling time is less than the look-back time at $z=1$ (roughly 7.7 Gyr; e.g., \citealt{2004ApJ...607..800B,2012AdAst2012E...6G}). By fitting the cooling time profile of ZwCl~235 with a power-law (see Fig. \ref{profile}), and locating the intersection with t$_{\text{cool}}=7.7$ Gyr we measured $r_{\text{cool}} = 78.5\pm1.4$ kpc. Then we extracted two spectra, one from the cooling region using an annulus centered on the BCG with inner radius 3$''$ and outer radius r$_{\text{out}}=$ r$_{\text{cool}}$, and the other extending from r$_{\text{cool}}$ to the edge of the chip to account for the ICM projected along the line of sight. The spectra were fitted with a \texttt{projct$\ast$tbabs$\ast$apec} model, which allowed to measure the properties of the ICM within r$_{\text{cool}}$: we found a temperature $kT = 2.74^{+0.06}_{-0.05}$ keV, a metallicity $Z=1.21\pm0.34$ and an electron density $n_{e}=(2.8\pm0.1)\times10^{-2}$ cm$^{-3}$ (with $\chi/$D.o.f = 274/265). The bolometric luminosity of the cooling region was measured to be $L_{bol}$[0.1 - 100 keV] $\equiv L_{\text{cool}}=(1.0\pm0.1)\times10^{44}$ erg s$^{-1}$. We can compare this value with the cooling luminosity of other systems in Tab. \ref{tab:objects}. Clusters at the top of the list have $L_{\text{cool}}\geq10^{45}$ erg s$^{-1}$ (e.g., Abell~1835, \citealt{mcnamara2006}, Abell~2204, \citealt{2009MNRAS.393...71S}), those closer to ZwCl~235 have $L_{\text{cool}}\geq10^{44}$ erg s$^{-1}$ (e.g., 2A0335+096, \citealt{2009MNRAS.396.1449S}), while those at lower H$\alpha$ luminosity have $L_{\text{cool}}\geq10^{43}$ erg s$^{-1}$ (e.g., Abell~1668, \citealt{pasini2021}, Abell~2495, \citealt{2019ApJ...885..111P}). This trend supports the link between the strength of ICM radiative losses and the amount of multi-phase gas at the cluster center (e.g., \citealt{2010ApJ...721.1262M,2020NatAs...4...10G}).
\begin{figure*}[ht]
\centering
\includegraphics[width=0.8\linewidth]{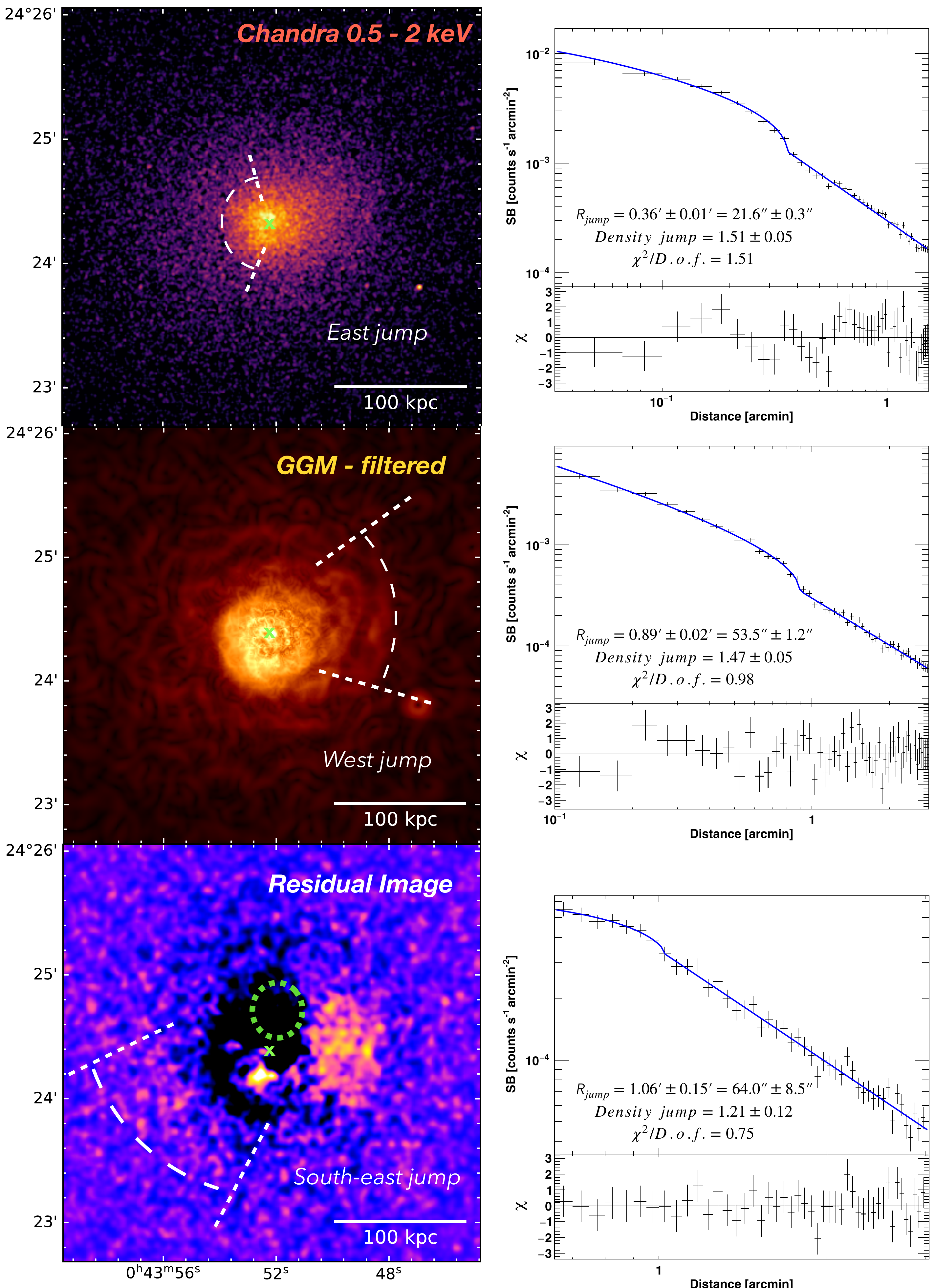}
\caption{\textit{Left panels:} from top to bottom, 0.5 - 2 keV band \textit{Chandra} image of ZwCl~235, GGM-filtered image, and double $\beta$-model residual image (see Sect \ref{fronts}). White dashed arcs mark the positions of the edges, white dashed lines show the opening angles of the sectors used for the spectral analysis, and the green cross indicates the position of the BCG. In the residual image, the green dashed ellipse indicates the X-ray depression identified by \citet{2016ApJS..227...31S} (see \cref{feedback}). \textit{Right panels:} from top to bottom, surface brightness profiles across the east, west and south-east edges. In each panel, the best-fit broken power-law model is overlaid in blue and detailed in the text, while the residuals are shown in the lower boxes.
              }
         \label{figfronts}
\end{figure*}
\subsection{Extended sloshing and cold fronts}
\label{fronts}
On large scales, the maps shown in Fig. \ref{figspectral} (\textit{upper panels}) reveal three bow-like discontinuities in temperature and entropy on opposite sides of the core, which seem nearly continuous in pressure. These properties are usually associated with cold fronts generated by sloshing of the cold ICM. In this context, \citet{2020MNRAS.496.2613B} first noted the spiral morphology of the ICM and suggested that sloshing could be present in ZwCl~235. \\ \indent In order to further investigate these features, we created a a Gaussian Gradient Magnitude (GGM) filtered image (see \citealt{2016MNRAS.460.1898S}) and a residual image. For the first one, we have filtered with the GGM filter the 0.5 - 7 keV image, using scales of 1, 2, 4, 6, 8 pixels; following the procedure outlined by \citet{2016MNRAS.457...82S} the resulting five images were combined using a weighted sum. The weights have been set to give smaller scale filters more contribution in the center, and larger scale filters more contribution in the outer regions. The final image is shown in Fig. \ref{figfronts}. For the second one, we subtracted the best fit double $\beta$-model described in the previous section from the original \textit{Chandra} image, obtaining the residual map shown in Fig. \ref{figfronts}.
\\ \indent These images highlight an excess to the west of the center, that is spatially connected to the eastern one forming a spiral-like feature. Moreover, we find a possible third edge south-east from the core, slightly visible also in the original image. Interestingly, the location of the three edges matches the position of the temperature discontinuities. To measure the magnitude of these jumps, surface brightness profiles were extracted from the 0.5 - 2 keV band \textit{Chandra} image, using sectors that encompass the discontinuities (see regions overlaid on Fig. \ref{figfronts}). The profiles were then fitted with a broken power-law model. The resulting surface brightness profiles and best fit models are shown in Fig. \ref{figfronts} (\textit{right panels}). We confirmed the presence of an eastern surface brightness edge at 21.6$''\pm$0.3$''$ ($\sim$34 kpc) from the center, characterized by a density jump of 1.51$\pm$0.05, and of a western surface brightness edge at 53.5$''\pm$1.2$''$ ($\sim$84 kpc) from the center, characterized by a density jump of 1.47$\pm$0.05. For the possible south-east edge we find a density jump of 1.21$\pm$0.12 at 64.0$''\pm$8.5$''$ ($\sim$100 kpc), but the relatively large uncertainties do not allow to securely detect it. \\ \indent To confirm the hypothesis of the cold front nature of the discontinuities, based on the spectral maps, we extracted the spectra of the ICM inside and outside each discontinuity, with a third region extending to the edge of the chip to account for deprojection. Fitting the 0.5 - 7 keV band spectra with a \texttt{projct$\ast$tbabs$\ast$apec} model returned the deprojected thermodynamic properties (temperature, metallicity, electron density, pressure and entropy) reported in Tab. \ref{tab:fronts}. The spectral analysis confirmed that each jump has the typical properties of cold fronts, in particular: 
\begin{itemize}
    \item the ICM temperature inside the east front is lower than outside ($kT_{\text{in}}/kT_{\text{out}} = 0.65\pm0.08$), and we measured a density jump of $n_{e,in}/n_{e,out} = 1.85\pm0.17$, and continuous pressure at the interface ($p_{\text{out}}/p_{\text{in}} = 1.2\pm0.2$);
    \item for the west front, we found lower ICM temperature ($kT_{\text{in}}/kT_{\text{out}} = 0.79\pm0.11$), higher density ($n_{e,in}/n_{e,out} = 2.39\pm0.22$) and continuous pressure ($p_{\text{out}}/p_{\text{in}} = 1.7\pm0.7$);
    \item for the south-east front, we found $kT_{\text{in}}/kT_{\text{out}} = 0.58\pm0.12$, $n_{e,in}/n_{e,out} = 2.22\pm0.19$ and $p_{\text{out}}/p_{\text{in}} = 1.3\pm0.5$.
\end{itemize}
Overall, our results support the hypothesis of \citet{2020MNRAS.496.2613B} that sloshing is shaping the X-ray morphology of ZwCl~235. 
\\ \indent Furthermore, we note that the spectral analysis of the cold fronts hints at the inner side having higher abundance than the outer side of the east and the west edges (see Tab. \ref{tab:fronts}). However, due to the relatively low number of counts, the abundance gradients are marginally significant  ($\sim$2.2$\sigma$ for the east front and $\sim$1.4$\sigma$ for the west front), and for the south-east front the metallicity cannot be constrained.
\begin{table*}[ht]
		\centering
		\caption{Spectral analysis of the three surface brightness discontinuities discussed in \cref{fronts}.}
		\renewcommand{\arraystretch}{1.5}
		\label{tab:fronts}
		\begin{tabular}{ccccccccc}
			\hline
			Discontinuity & R$_{\text{i}}$ & R$_{\text{o}}$ & Counts & Z & $kT$ &   $n_{\text{e}}$ &$p_{\text{ICM}}$ & $K_{\text{ICM}}$\\
			& [kpc] &[kpc] & & [Z$_{\odot}$] & [keV] &  [10$^{-3}$ cm$^{-3}$] & [10$^{-11}$ erg cm$^{-3}$] & [keV cm$^{2}$]\\
			\hline
			East (34 kpc) & 19 & 34 &2436 (99.4 \%) &1.01$^{+0.29}_{-0.29}$ &2.34$^{+0.20}_{-0.21}$  &  30.5$^{+1.8}_{-1.4}$& 20.9$^{+2.1}_{-1.9}$ & 23.9$^{+2.2}_{-1.6}$\\
			& 34 & 59  & 2389 (97.9 \%) & 0.40$^{+0.17}_{-0.17}$&3.56$^{+0.28}_{-0.26}$  &  16.5$^{+0.4}_{-0.4}$& 17.3$^{+1.5}_{-1.4}$ & 54.8$^{+4.4}_{-4.3}$\\
			& 59 & 390 & 11319 (77.6 \%) & 0.57$^{+0.09}_{-0.10}$  & 3.98$^{+0.15}_{-0.15}$  &  1.7$^{+0.1}_{-0.1}$& 2.0$^{+0.2}_{-0.2}$ & 274.6$^{+20.1}_{-20.1}$ \\
			\hline
			West (84 kpc) & 47 & 84 &2724 (97.7 \%) &0.88$^{+0.24}_{-0.24}$& 2.72$^{+0.19}_{-0.20}$  & 11.3$^{+0.4}_{-0.4}$& 9.0$^{+0.7}_{-0.8}$ & 53.9$^{+4.2}_{-4.2}$\\
			& 84 & 142  &2249 (92.5 \%) &0.41$^{+0.28}_{-0.29}$& 3.43$^{+0.51}_{-0.45}$   & 4.6$^{+0.2}_{-0.2}$&4.9$^{+0.7}_{-0.7}$ &123.5$^{+16.7}_{-16.7}$\\
			& 142 & 265 & 2833 (81.1 \%) & 0.37$^{+0.17}_{-0.18}$ & 4.12$^{+0.33}_{-0.30}$  & 2.5$^{+0.1}_{-0.1}$&3.0$^{+0.2}_{-0.3}$ &224.9$^{+18.5}_{-21.3}$\\
			\hline
			South-east (100 kpc) & 59 & 100  &1429 (96.3 \%) &$ < 0.43$& 2.45$^{+0.34}_{-0.34}$  &  8.7$^{+0.4}_{-0.4}$& 6.2$^{+0.9}_{-0.9}$ & 58.0$^{+8.3}_{-8.3}$\\
			& 100 & 172 &1565 (89.9 \%) &$<$0.85& 4.20$^{+0.45}_{-0.44}$  &  3.9$^{+0.2}_{-0.2}$& 4.8$^{+0.6}_{-0.6}$ & 168.5$^{+18.6}_{-18.6}$\\
			& 172  & 515 & 2454 (54.9 \%) & 0.51$^{+0.25}_{-0.33}$  & 4.33$^{+0.51}_{-0.50}$  &  0.9$^{+0.1}_{-0.1}$& 1.2$^{+0.1}_{-0.1}$ & 459.7$^{+54.7}_{-45.6}$ \\
			\hline
		\end{tabular}
		\tablefoot{(1) name of the discontinuity and distance from the center; (2-3) inner and outer radius of the sector used for the spectral analysis (the first sector encloses the front, the second sector contains the region outside the front, and the third, outermost sector is used for deprojection); (4) net photon counts (fraction w.r.t. the total counts); (5) Abundance; (6) temperature; (7) electron density; (8) pressure of the ICM; (9) entropy of the ICM. The $\chi^{2}/d.o.f.$ is 407.8/383 (1.07) for the east front, 231.5/252 (0.92) for the west front and 196.2/243 (0.81) for the south-east front.}
\end{table*}	
Nonetheless, the presence of higher metallicity inside the cold fronts would be consistent with the sloshing scenario (see e.g., \citealt{2013AN....334..422G}). As the cold, enriched central gas oscillates around the center, it comes at contact with the hotter and less abundant outer ICM, thus generating not only temperature and density, but also abundance gradients. While this is only hinted at in ZwCl~235, enhanced metallicity along low temperature spirals are common (e.g., among the systems in Tab. \ref{tab:objects}, the cluster 2A0335+096 shows a low temperature, high metallicity swirl wrapped around the core, \citealt{2009MNRAS.396.1449S}).

\subsection{Evidence for AGN feedback in ZwCl~235}
\label{feedback}
In their systematic search of cavities on a large cluster sample, \citet{2016ApJS..227...31S} found a depression in the residual image of ZwCl~235, located approximately 14 kpc north-west of the center, that was tentatively classified as an AGN-inflated X-ray cavity. On the other hand, from the LOFAR observation of the radio galaxy at the center of ZwCl~235, \citet{2020MNRAS.496.2613B} noted that the 144 MHz lobes are orthogonal to the X-ray cavity identified by \citet{2016ApJS..227...31S}, thus questioning its classification. Our aim is to get a closer, dedicated look at the activity of the AGN in ZwCl~235. In this section we investigate the properties and morphology of the central radio galaxy, and we search for imprints of feedback (mainly in the form of X-ray cavities) in the innermost region of the cluster. 
\begin{figure*}[ht]
\includegraphics[width=1\linewidth]{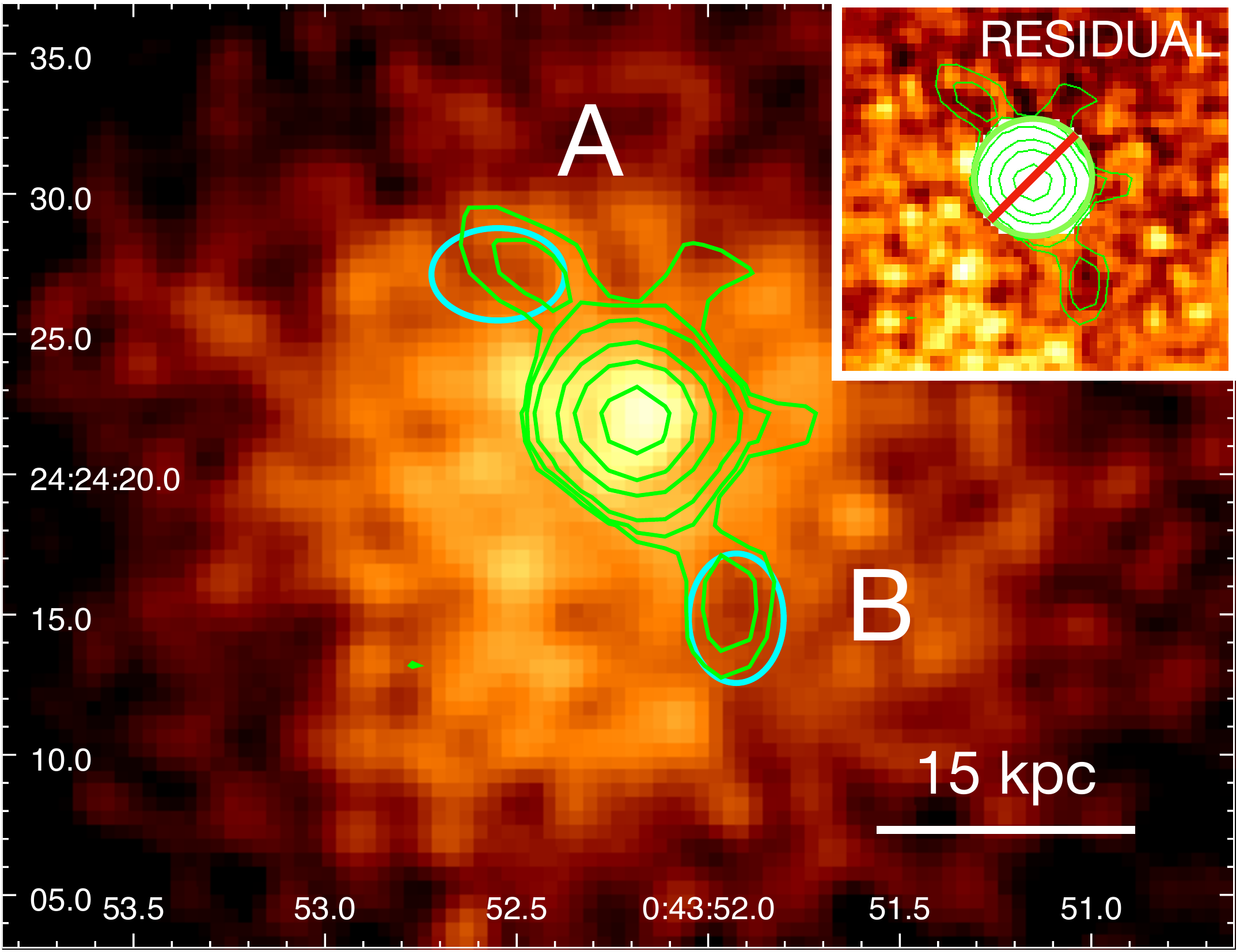}
\caption{\textit{Main panel}: \textit{Chandra} image (0.5 - 7 keV) of the galaxy cluster core, smoothed with a Gaussian of kernel size 1.5$''$. The overlaid green contours represent the emission of the central radio galaxy at 3 GHz (same as in Fig. \ref{figcav_paper}), while the cyan ellipses show the position of the putative X-ray cavities discussed in \cref{feedback}. \textit{Upper zoom-in}: Double $\beta$-model residual image (see Sect. \ref{fronts}), showing depressions at the position of the NE - SW radio lobes (contours match those of the main panel). The white barred central region is caused by the excision of the inner 3$''$ from the image during fitting. 
              }
         \label{feedback_fig}
\end{figure*}
\\ \indent  Fig. \ref{feedback_fig} shows the VLASS 3 GHz radio contours of the radio galaxy at the center of ZwCl~235, superimposed on the \textit{Chandra} 0.5 - 7 keV image and on the double $\beta$-model residual image. Guided by the radio morphology, we found evidence for the presence of two depressions that match the orientation and extension of the 3 GHz radio lobes (see green ellipses in Fig. \ref{feedback_fig}). The depressions have diameters of $7.5\times5.3$ kpc, and are located at a distance of $\sim$12 kpc from the center. By comparing the surface brightness of the cavities with that of the surrounding ICM at the same distance from the center, we measured a deficit of 20$\%$ at a significance of $\sim$2.3$\sigma$ (for comparison, we determined that other regions of apparently low surface brightness within $\sim$20 kpc from the center  represent depressions of at most $\sim$5\% w.r.t the surrounding ICM). While a deeper \textit{Chandra} exposure would be required to securely detect the two features, the comparison with the morphology of the radio galaxy supports their interpretation as real cavities. We list in Tab. \ref{tab:cavities} the physical properties of the two depressions. 
\\ \indent To compare the depressions we identified with the putative cavity reported in \citet{2016ApJS..227...31S} we consider the relation between the cavity area (\textit{a}) and distance from the center (\textit{D}) discussed in their work. Such relation ($\log{a} = (0.97\pm0.02)\log{D}-0.15\pm0.02$) would predict an area for the cavities in ZwCl~235 identified here of $\sim$130 kpc$^{2}$. Thus, while within the scatter found in \citet{2016ApJS..227...31S}, the putative cavities in ZwCl~235 are smaller than expected, having an observed area of $\sim$32 kpc$^{2}$. Nonetheless, even the putative cavity identified by \citet{2016ApJS..227...31S} has a larger area than expected (533 kpc$^{2}$). In this respect, considering the residual image of ZwCl~235, we argue that the location of the putative cavity identified by \citet{2016ApJS..227...31S} in their residual image (see green ellipse in Fig. \ref{figfronts}) suggests that the depression is a consequence of the sloshing pattern of the ICM of ZwCl~235, rather than an AGN-inflated bubble. Indeed, residual images of sloshing clusters can show negative depressions due to the spiral morphology of the ICM (see e.g., \citealt{2020ApJ...892..100U} for a discussion of such features); in cases where X-ray cavities are also present, the interpretation of the residual patterns requires a detailed analysis (see e.g., the analysis of NGC~1550 in \citealt{2020MNRAS.496.1471K}, or of 2A0335+096, \citealt{kokotanekov2018}). In the work of \citet{2016ApJS..227...31S}, this issue is highlighted for Abell~1991, where a prominent cold front causes over-subtraction of the surface brightness, generating a large depression that resembles a cavity.
\\ \indent  In order to determine the history of AGN feedback in this galaxy cluster, we derived the age of the putative cavities combining classical techniques: the sound speed age, the refill time, and the buoyancy age of the cavities (see e.g., \citealt{2004ApJ...607..800B}). The sound speed age was derived by using the temperature of the ICM at the distance of the cavities from the center (8$''$), measured from the azimuthal profile of Fig. \ref{profile}. The acceleration at the cavity position (\textit{g}$\approx$9.8$\times10^{-8}$ cm s$^{-2}$), needed to measure the refill and buoyancy ages, was determined from the mass profile of ZwCl~235 obtained by \citet{2018ApJ...853..177P}. From the timescales reported in Tab. \ref{tab:cavities}, it is clear that the three methods are in good agreement with each other, supporting an average age for the AGN outburst of $\approx$17 Myr. This timescale is consistent with the upper limit of $\lessapprox70$ Myr for the radiative age of the synchrotron emitting particles (see \cref{radiogalaxy}). 
\begin{table}[ht]
	\centering
	\renewcommand{\arraystretch}{1.5}
	\caption{Derived properties of the X-ray depressions discussed in \cref{feedback} if interpreted as cavities.}
	\label{tab:cavities}
	\begin{tabular}{l|c|c}
		\hline
		  & Cavity A & Cavity B \\
		\hline
		\hline
		 Major semi-axis [kpc]       & 3.8 & 3.8 \\
		 Minor semi-axis [kpc]       & 2.7 & 2.7 \\
		 Distance from the AGN [kpc] & 11.5 & 12.5 \\
		 Area [kpc$^{2}$]   & 32.2 & 32.2 \\
		 \hline
		 Sound speed age [Myr] & $15.5\pm1.8$ & $16.5\pm1.9$ \\
		 Buoyancy age [Myr]    & $16.5\pm1.4$ & $17.6\pm2.9$ \\
		 Refill age [Myr]      & $19.9\pm1.4$ & $19.9\pm3.1$ \\
		 \hline
		 ICM Pressure [10$^{-10}$ erg cm$^{-3}]$ & $2.3\pm0.2$ & $2.3\pm0.2$ \\
		 Cavity Volume [kpc$^{3}$]               & $131\pm13$  & $131\pm13$  \\
		 Cavity Power [10$^{42}$ erg s$^{-1}$]   & $6.0\pm0.7$ & $6.0\pm0.7$ \\
		 \hline
	\end{tabular}
\end{table}
\\ \indent The mechanical power required to inflate X-ray cavities is typically measured as $4pV/t_{\text{age}}$ (e.g., \citealt{2004ApJ...607..800B}). From the azimuthal pressure profile (see Fig. \ref{profile}) we determined the pressure $p$ of the ICM at the distance of the cavities from the center. The volume $V$ of the cavities was computed assuming that the depressions can be approximated as oblate ellipsoids (a relative uncertainty of 10\% was associated with volumes). For each cavity, the last three rows of Tab. \ref{tab:cavities} report the ICM pressure, volume and power (using an average $t_{\text{age}}\sim17$ Myr). We measured a total cavity power of $(1.2\pm0.2)\times10^{43}$ erg s$^{-1}$, which is similar to the cavity power measured in Abell~1668 \citep{pasini2021}, Abell~2495 \citep{2019ApJ...885..111P} and Abell~2052 \citep{blanton2011}, that are close in H$\alpha$ luminosity and X-ray flux to ZwCl~235 (see Tab. \ref{tab:objects}). Such cavity power is typical of mild AGN outbursts in galaxy clusters, whereas more violent episodes of AGN activity can lead to outburst powers of 10$^{45}$ erg s$^{-1}$ (considering the comparator clusters in Tab. \ref{tab:objects} as e.g., Abell~1835, \citealt{mcnamara2006}, and Abell~2204, \citealt{2009MNRAS.393...71S}), up to $10^{46}$ erg s$^{-1}$ (see e.g., \citealt{2014MNRAS.442.3192V}).

\subsection{Metallicity of the ICM}
\label{metal}
\begin{figure}
\includegraphics[width=1\linewidth]{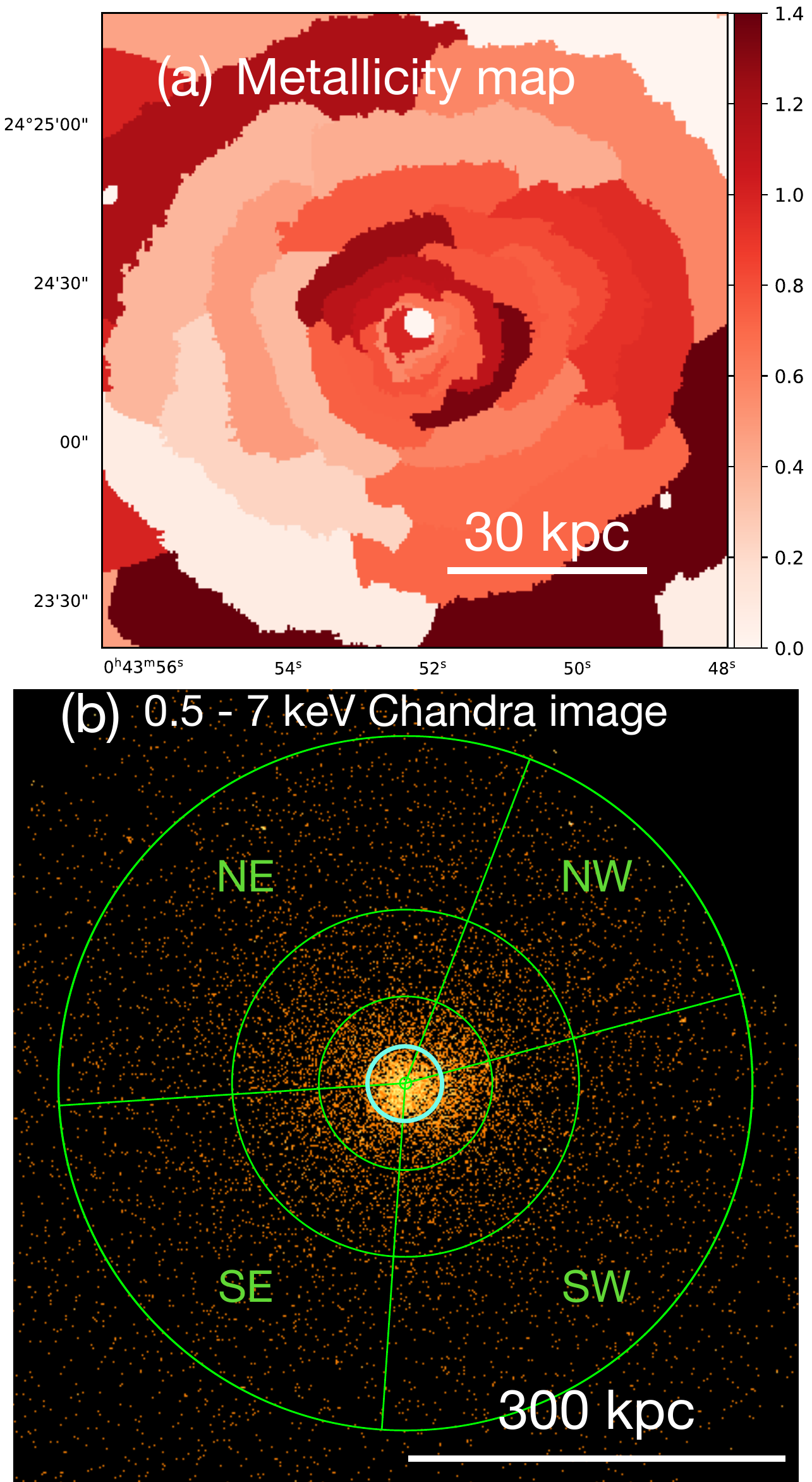}
\caption{\textit{Panel a:} Metallicity map of ZwCl~235 (see \cref{general} for details on how the map was built), showing two metal rich arcs in the NE - SW direction at roughly 20 kpc from the center. The relative error is $\sim$30\%. \textit{Panel b:} 0.5 - 7 keV raw \textit{Chandra} image of ZwCl~235. The green sectors correspond to the regions used to study the azimuthal and radial variation of abundances in the ICM (see \cref{metal} and Tab. \ref{tab:metal}). The metal-rich arcs visible in \textit{panel a} are located within the innermost cyan circle.
              }
         \label{figZ}
\end{figure}

By examining the radial profile of metallicity shown in Fig. \ref{profile}, we do not find evidence for a radial dependence of the abundance in ZwCl~235, i.e. within the large uncertainties our measurements are consistent with a flat profile. While cool core clusters are known to show central enhancements in iron abundance (e.g., \citealt{2004A&A...416L..21B}; for instance, this applies to Abell~1068, \citealt{2004ApJ...601..184W}, Abell~1835, \citealt{mcnamara2006}, or Abell~2204, \citealt{2009MNRAS.393...71S}), ZwCl~235 would not be unique in having a flat profile within a few 100s kpc from the center (see e.g., Abell~1795, \citealt{kokotanekov2018}, and Abell~1991, \citealt{sharma2004}).
The highest metallicity is found in the third bin of the radial profile, that corresponds to the range 12$''$-16$''$. Within this radial range, in the map of abundances in ZwCl~235 (shown in Fig. \ref{figZ}, \textit{panel a}) there are two higher abundance arc-like regions north-east and south west of the center, at a distance of $\sim$20$''$ (roughly 30 kpc), with a position angle of approximately 132$^{\circ}$.
However, given its relatively large error, the metallicity map alone should be used only as a guide to identify interesting features for a further, detailed analysis. To properly investigate metallicity gradients, we then measured the metallicity radial profile in four sectors encompassing the enriched arc-like regions identified by the spectral map (see Fig. \ref{figZ}). The spectrum of each region was fitted with a \texttt{tbabs$\ast$apec} model; the best-fit metallicity values are shown in Tab. \ref{tab:metal}. 
\\ \indent  With these improved statistics we recover a trend of decreasing metallicity going far from the center for the north-east (NE), south-east (SE) and south-west (SW) sectors, as expected in a cool core cluster. Instead, the highest metallicity in the north-west sector is found within the second annulus. On the one hand, the uncertainties are still large enough to account for this difference; on the other hand, this could be explained by considering that this region overlaps with the inner side of the west front, for which we found hints of a metallicity gradient.
Moreover, within 30 kpc the metallicity is higher along the NE and SW sector, that include the two metal-rich arcs of the metallicity map. The azimuthal gradient is more pronounced with respect to the NW sector, since the opposite SE sector matches the position of the east front. Beyond 30 kpc, the four sectors have similar metallicities. 
\\ \indent  While the uncertainties prevent us from definitely attesting the existence of metal rich arcs (robust detection of abundance gradients require deep \textit{Chandra} data, see \citealt{2015MNRAS.452.4361K}), the consistent indications from the metallicity map and the study in sectors support the presence of two metal-rich regions between $5-30$ kpc from the center, with a position angle of $\sim$132$^{\circ}$. Assuming that the metal-rich arcs are real, we discuss the possible origins of this abundance distribution in \cref{metaldist}. 
\begin{table}[ht]
	\centering
	\renewcommand{\arraystretch}{1.5}
	\caption{Azimuthal and radial variation of metallicity in the ICM (see \cref{metal} and Fig. \ref{figZ}, \textit{panel b}).
	}
	\label{tab:metal}
	\begin{tabular}{|c||c|>{\columncolor[gray]{0.8}}c|c|>{\columncolor[gray]{0.8}}c|}
		\hline
		 R$_{\text{in}}-$R$_{\text{out}}$ & Z$_{NW}$ & Z$_{NE}$ & Z$_{SE}$ &Z$_{SW}$ \\
		 $[\text{kpc}]$ & [Z$_{\odot}$] & [Z$_{\odot}$] & [Z$_{\odot}$] & [Z$_{\odot}$] \\
		\hline
		\hline
		 4.7 - 30 & $0.58^{+0.29}_{-0.31}$ & $1.11^{+0.22}_{-0.20}$ & $0.75^{+0.15}_{-0.14}$ & $0.99^{+0.17}_{-0.16}$ \\
		 \hline
		 30 - 70 & $0.72^{+0.24}_{-0.24}$ & $0.56^{+0.18}_{-0.17}$ & $0.56^{+0.19}_{-0.21}$ & $0.81^{+0.16}_{-0.17}$\\
		 \hline
		 70 - 140 & $0.53^{+0.25}_{-0.22}$ & $0.64^{+0.18}_{-0.19}$ & $0.43^{+0.15}_{-0.17}$ & $0.57^{+0.18}_{-0.18}$\\
		 \hline
		 140 - 280 & $0.31^{+0.24}_{-0.15}$ & $0.48^{+0.16}_{-0.14}$ & $0.50^{+0.25}_{-0.23}$ & $0.33^{+0.12}_{-0.11}$\\
		 \hline
	\end{tabular}
	\tablefoot{(1) inner and outer radius of the annular sector; (2) abundance in the north-west sector; (3) abundance in the north-east sector; (4) abundance in the south-east sector; (5) abundance in the south-west sector. The grey columns highlight the sectors that encompass the metal-rich arcs (see Fig. \ref{figZ}, \textit{panel a}), which are found in the first radial bin (i.e. within 30 kpc from the center). }
\end{table}

\subsection{The nature of the X-ray emission from the inner 3$''$}
\label{bcg}
\begin{figure}[!ht]
\includegraphics[width=1\linewidth]{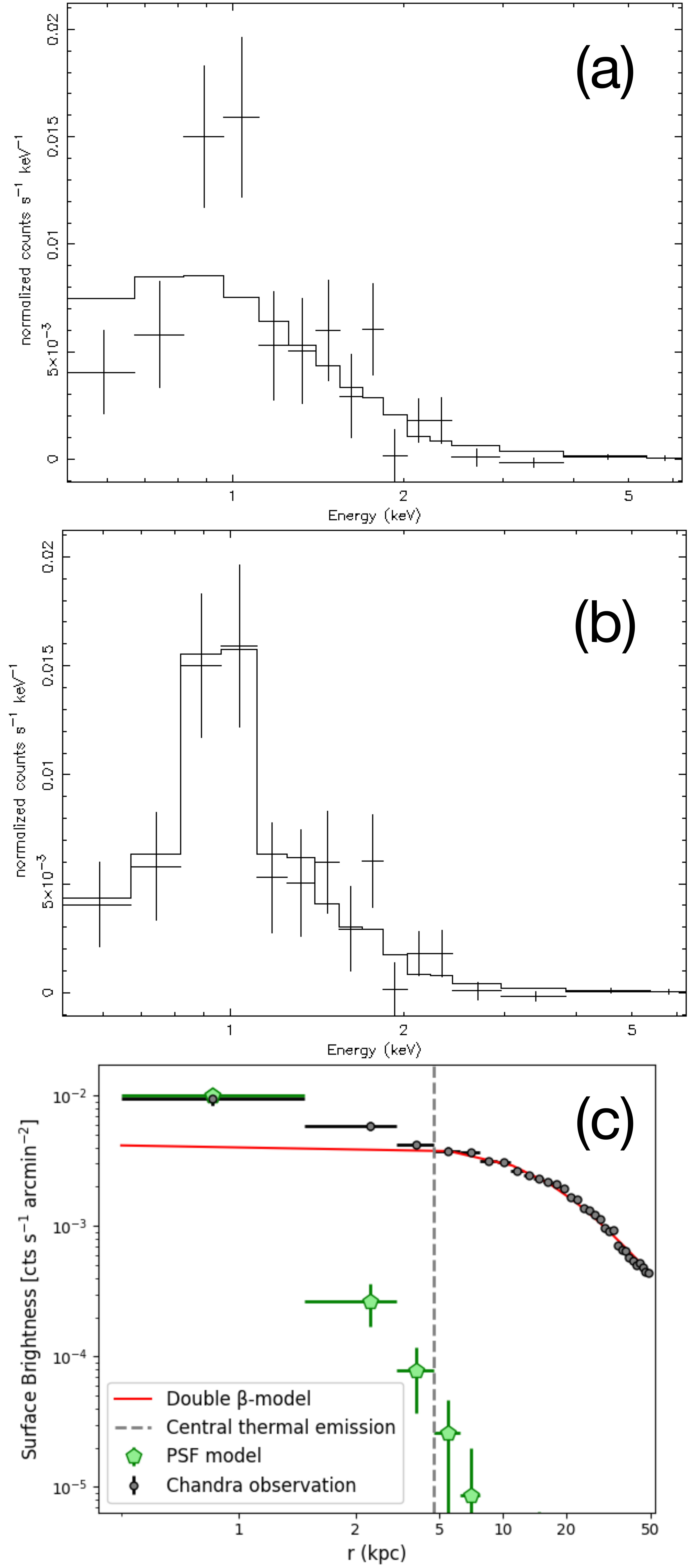}
\caption{\textit{Panel a:} 0.5 - 7 keV \textit{Chandra} spectrum of the BCG (black crosses), with the absorbed power-law model (first row of Tab. \ref{tab:bcgspectrum}) over-plotted as a black line. \textit{Panel b:} Same as in \textit{panel a}, with the absorbed thermal model (second row of Tab. \ref{tab:bcgspectrum}) over-plotted as a black line. In both panels, the data are binned to 10 cts bin$^{-1}$ for plotting purposes. \textit{Panel c:} Comparison between the 0.5 - 2 keV surface brightness profile of the PSF (green points) and that of the \textit{Chandra} observation of ZwCl~235 in the inner 50 kpc. The grey dashed line marks the size of the extraction region of the BCG spectrum (3$''\sim$ 4.7 kpc; see \cref{bcg}).
              }
         \label{bcgpsf}
\end{figure}

The \textit{Chandra} image of Fig. \ref{figchandra} reveals a bright spot in the inner 3$''$ at the center of the cluster. Considering the VLBA detection of the AGN core (see \cref{radiogalaxy}), and the co-spatiality of the X-ray bright spot with the radio core as seen at 3 GHz, we need to consider if this X-ray emission is non-thermal. By producing images in different energy bands (0.5-2.0, 2.0-4.0, 4.0-7.0 keV) we verified that the emission fades into the background above 2-3 keV. Thus, if the source is described by a power-law model (emitting non thermal X-ray radiation), the spectral index could be very steep. Alternatively, the emission could arise from a thermal component with a low (1-2 keV) temperature. To discern between the two models, we extracted the source spectrum from the inner 3$''$, and we extracted the local background spectrum from an annulus extending from 3$''$ to 8$''$ from the center, in order to subtract from the source spectrum the contribution of the surrounding ICM. The spectrum was fit with an absorbed power-law (\texttt{tbabs$\ast$po}), leaving the spectral index $\Gamma$ and the normalization free to vary. We adopted the $C$-statistics \citep{1979ApJ...228..939C} to account for the relatively low number of counts (542 counts, including background). The resulting spectral index of $\Gamma=2.36^{+0.22}_{-0.20}$ is steep (the typical $\Gamma$ found in BCGs lie in the range 1 - 2, see e.g., \citealt{2013MNRAS.431.1638H,2018ApJ...859...65Y}), as expected from the images produced in different energy bands. The $C/d.o.f.$ is 174/151. We verified that the inclusion of an intrinsic absorber (\texttt{ztbabs}) is not required. 
\\ \indent While this fit constrains quite well the power-law index, there are strong positive residuals (at more than $5\sigma$) in the 0.8-1.0 keV spectral window (see Fig. \ref{bcgpsf}, panel \textit{a}). According to e.g., \citet{2007ApJ...657..197S}, the presence of a soft bump in the X-ray spectrum of brightest cluster galaxies is caused by a thermal emission with temperature of $\approx$1 keV coming from the galactic thermal halo, generating the rise in the region of the Fe-L line. In order to test this hypothesis, we performed a fit of the spectrum with a \texttt{tbabs$\ast$apec} model, leaving the temperature, abundance and normalization free to vary. We find an improvement in the fit w.r.t. the power-law ($C/d.o.f.$=153/150) for a thermal plasma with $kT=1.45^{+0.18}_{-0.15}$ keV and $Z=0.75^{+0.45}_{-0.40}\,Z_{\odot}$. The residuals around 0.8-1.0 keV are accounted for by this model (\textit{panel b} of Fig. \ref{bcgpsf}). By fitting a combined \texttt{tbabs$\ast$(apec+po)} model the normalization of the power-law approaches zero and the spectral index $\Gamma$ assumes unphysical values ($\sim$-3), which suggests that the power-law component is not required. To constrain the contribution of any non-thermal emission, we fixed the power-law index to $\Gamma=1.9$ (following e.g., \citealt{2013MNRAS.432..530R,2013MNRAS.431.1638H}), obtaining a 1$\sigma$-confidence upper limit on the 2 - 10 keV flux of the power-law of $f_{2-10} \lessapprox 2.0\times10^{-14}$ erg cm$^{-2}$ s$^{-1}$ (see third row of Tab. \ref{tab:bcgspectrum}). 
Therefore, we conclude that the X-ray spectrum of the inner 3$''$ arises mainly from a kernel of $\sim$1 keV thermal plasma, rather than from an unresolved power-law. Deeper X-ray observations might detect non-thermal emission coming from the BCG, which would be expected given the VLBA detection of the AGN and its flat spectral index at GHz frequencies (see also \citealt{2015MNRAS.453.1201H}).
\\ \indent As a sanity check, we considered the effect of using the blank-sky to model the background, which provided consistent results with those presented here (the details of the blank-sky test are presented in the Appendix \ref{altbcg}). Considering this, and the fact that the use of a local background is the standard method used to study the X-ray emission from thermal plasma at 1-2 keV in the central kpc of clusters (see e.g., \citealt{2007ApJ...657..197S}), we selected the thermal model described in Tab. \ref{tab:bcgspectrum} as our best-fit model for the central X-ray emission. 
\\ \indent Since our analysis suggests that the spectrum within 3$''$ from the center is likely of thermal origin, the emission should appear extended when compared to the PSF (see e.g., \citealt{2007ApJ...657..197S}). Using the Chandra ray-tracing program ChaRT \citep{2003ASPC..295..477C} we generated a model of the PSF between 0.5-2.0 keV. Then, we extracted a surface brightness profile of bin width 1$''$ from the \textit{Chandra} image (in the 0.5-2.0 keV band) and from the PSF model. The profiles are compared in Fig. \ref{bcgpsf}. The surface brightness profile for $r<3''(\sim5$ kpc) is in excess at more than 5$\sigma$ with respect to the PSF profile, indicating that the emission is extended. Moreover, the X-ray emission in the inner 5 kpc is also in excess of the double $\beta$-model describing the ICM surface brightness profile outside this radius (red line in Fig. \ref{bcgpsf}; see \cref{general}). Therefore, our analysis unveiled that the bright soft X-ray emission from the inner $\approx$5 kpc originates from thermal gas at $\sim$1 keV, whose surface brightness exceeds the expectations from the inward extrapolation of the azimuthally-averaged ICM surface brightness profile. 
Since the source is extended, and the use of a local background removed the contribution of the ICM projected along the line of sight, it is possible to derive the electron density n$_{e}$ of this thermal component from the normalization of the \texttt{apec} component (assuming a volume $V = 4/3\pi r^{3}$, with $r=4.7$ kpc). Combining the density and temperature we derived the pressure, entropy and cooling time. We measured n$_{e}=0.10\pm0.01$ cm$^{-3}$, p$=4.4\pm0.5\times10^{-10}$ erg cm$^{-3}$, K$=6.6\pm0.8$ keV cm$^{2}$ and $t_{\text{cool}}=238\pm34$ Myr. To facilitate a comparison with the results on the ICM at $r > 3''$, we include the properties of the thermal gas within 3$''$ on the plots of Fig. \ref{profile} (gray star). As a further validation of these results, we tested fitting the spectrum of the inner 3$''$ together with the spectra of the annuli between 3$''$-180$''$ (see \cref{general}) with a \texttt{projct$\ast$tbabs$\ast$apec} model, finding good agreement with the values reported here and shown in the plot of Fig. \ref{profile}. We observe that the pressure within 4.7 kpc matches that of the surrounding ICM ($p_{ICM} = 4.5\pm0.7\times10^{-10}$ erg cm$^{-3}$, see Fig. \ref{profile}). 
\\ \indent Since ZwCl~235 is a cool core cluster (see \cref{general}), it could be that this central thermal component represents the coldest, densest phase of the ICM; in particular, cooling of the central gas, which is more efficient at progressively decreasing radii, could be the origin of the 1 keV thermal plasma found on top of the BCG. Indeed, many other galaxy clusters show the co-spatiality of soft thermal emission with the region where a non-thermal X-ray point source is found or expected (see e.g., \citealt{2013MNRAS.432..530R}). On the other hand, this central soft component resembles the thermal \say{coronae} of BCGs discussed in \citet{2007ApJ...657..197S} and in \citet{2009ApJ...704.1586S}, i.e. X-ray emitting halos of gas surrounding the central elliptical galaxy of the cluster with temperature, entropy, density and cooling time significantly different from those of their ambient ICM (see also \citealt{2009ApJS..182...12C}). In this respect, the temperature, entropy, and cooling time of the central thermal component in ZwCl~235 are a factor of $\sim$2, $\sim$3 and $\sim$3 lower than those of the ICM outside 3$''$, respectively. Additionally, the total gas mass of the 1.4 keV plasma is $\approx$10$^{9}$ M$_{\odot}$, which is in agreement with the mass estimates for BCGs' thermal coronae of \citet{2007ApJ...657..197S}. Among the systems listed in Tab. \ref{tab:objects}, a thermal corona has been found surrounding the BCG of Abell~2634 \citep{2007ApJ...657..197S}. Considering the above indications, in the following we assume that the 1.4 keV thermal kernel represents the BCG thermal corona. However, we caution that the \say{coronal} scenario would not indicate the existence of two disconnected gas components at the cluster core (the corona and the ICM) spatially separated by a well-defined boundary. Rather, multi-temperature gas phases are likely to co-exist in the inner 10 kpc, with the thermal gas at progressively smaller radii being gradually more of stellar origin (see e.g., \citealt{1999ApJ...512...65B}). Refining the bin width of radial profiles on scales of e.g., 1$''$ would reveal if there is actually a smooth decrease in temperature or a sudden drop, but the number of counts in the available \textit{Chandra} data are insufficient.
\begin{table*}[ht]
	\centering
	\renewcommand{\arraystretch}{1.3}
	\caption{Spectral analysis of the X-ray spectrum of the BCG using a local background (see \cref{bcg} for details).
	}
	\label{tab:bcgspectrum}
	\begin{tabular}{l|cc|ccc|c}
		\hline
		 Model & $\Gamma$ & $norm_{pl}$ &$kT$ &Z&$norm_{\text{th}}$ & $C/d.o.f.$ \\
		  &  & [ph keV$^{-1}$ cm$^{-2}$ s$^{-1}$] &[keV] &[Z$_{\odot}$] & [ph keV$^{-1}$ cm$^{-2}$ s$^{-1}$] &  \\
		\hline
		    Non-thermal (\texttt{tbabs$\ast$po})& $2.36^{+0.22}_{-0.20}$ & $1.80^{+0.23}_{-0.22}\times10^{-5} $& (...)  & (...) & (...) & 174/151\\
		    Thermal (\texttt{tbabs$\ast$apec}) & (...)  & (...) & $1.45^{+0.18}_{-0.15}$& $0.75^{+0.45}_{-0.40}$ & $5.57^{+0.80}_{-0.95}\times10^{-5}$ & 153/150\\
		    Combined (\texttt{tbabs$\ast$(apec+po)})& 1.9 (Fixed) & $<5\times10^{-6}$ & $1.40^{+0.19}_{-0.16}$& $0.65^{+0.51}_{-0.43}$ & $5.33^{+0.92}_{-0.93}\times10^{-5}$ & 155/148\\
		  \hline
	\end{tabular}
	   \tablefoot{(1) model used to fit the spectrum; (2) spectral index of the power-law; (3) power-law normalization; (4) temperature of the thermal model; (5) abundance of the thermal model; (6) \texttt{apec} normalization; (7) $C$-statistics/degrees of freedom.}
\end{table*}	

\subsection{The bright X-ray filament}
\label{brightf}
\begin{figure}
\includegraphics[width=1\linewidth]{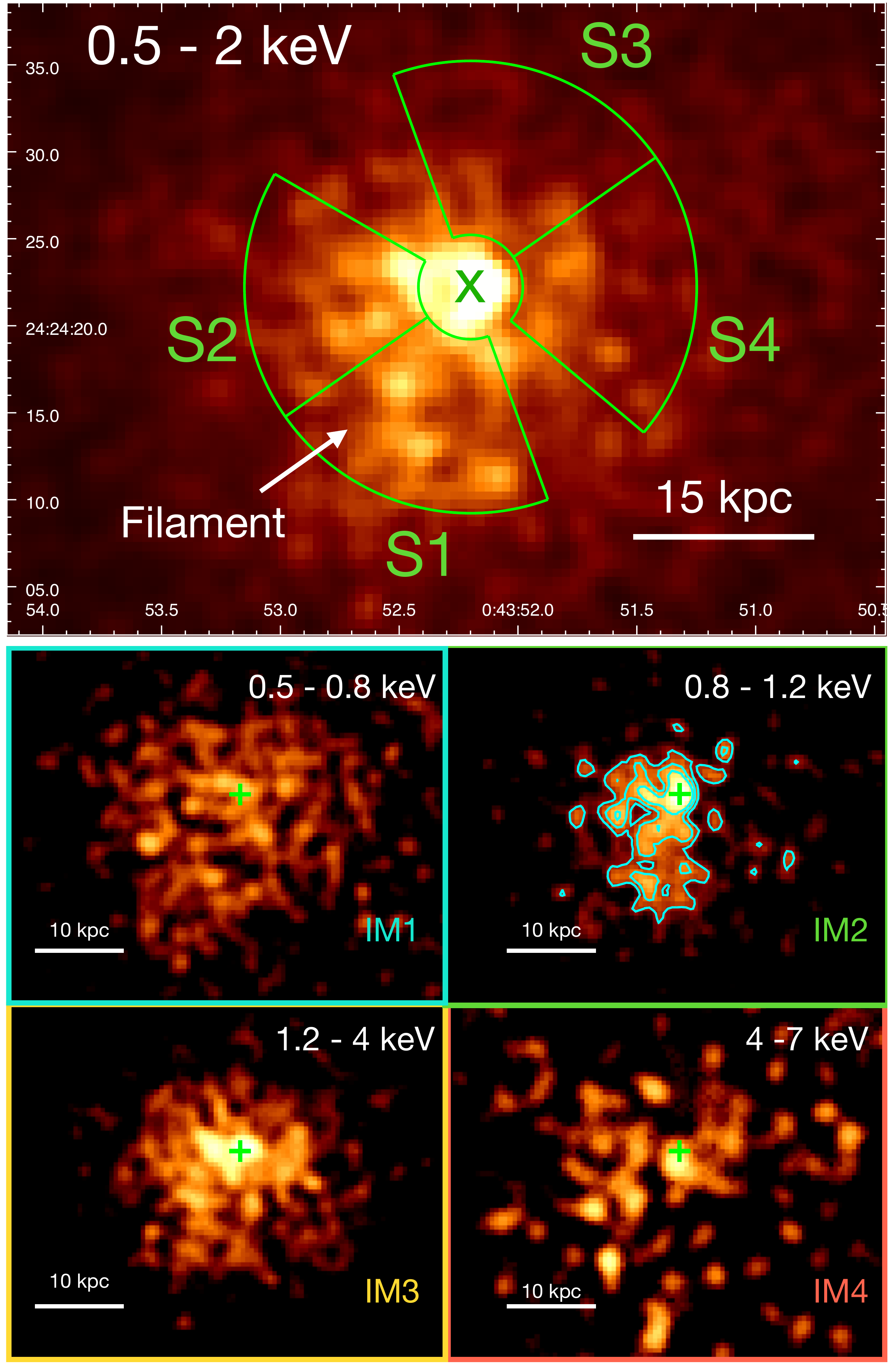}
\caption{\textit{Upper panel:} \textit{Chandra} image (0.5 - 2 keV band) of ZwCl~235, smoothed with a Gaussian of kernel size 1.5$''$, with the four sectors used to study the properties of the filament, that is encompassed by sector S1, overlaid in green (see \cref{brightf}). The green cross marks the center of the BCG. \textit{Lower panels:} background subtracted, exposure corrected \textit{Chandra} images of the cluster core in different energy bands, used to show that the $\sim$1 keV thermal plasma is found predominantly along the filament. The images are smoothed with a Gaussian of kernel size 1.5$''$ and have matching color scales. The cyan contours in IM2 are  spaced by a factor of $\sqrt{2}$, with the highest being 7$\times$10$^{-8}$ cts s$^{-1}$ cm$^{-2}$.}
         \label{figband}
\end{figure}

The inspection of the \textit{Chandra} image of ZwCl~235 revealed the presence of a $\sim$20 kpc-long bright filament in the central regions of the ICM (see \cref{general} and Fig. \ref{figchandra}). The filament, apparently connected to the central kernel of thermal plasma at 1.4 keV, extends to the south direction along the border of the SW radio lobe. In this section we perform a spectral analysis of the X-ray emission from the filament, with the aim of comparing the spectral properties of the bright feature with those of the surrounding ICM. 
\\ \indent In order to properly account for the asymmetrical thermodynamic distribution caused by sloshing, we extracted and fitted the spectrum of four sectors centered on the BCG with inner radius 3$''$ and outer radius 13$''$ (see \textit{upper panel} in Fig. \ref{figband}). The width of the sectors has been chosen to: (1) encompass the filament (region S1), (2) avoid the cavity region, (3) have two sectors east and west of the nucleus (in order to account for the asymmetry in thermodynamic properties induced by sloshing). Each spectrum was first fit with a one-temperature (1T) model (\texttt{tbabs$\ast$apec}), with the temperature, normalization and abundance left free to vary. From the resulting values reported in Tab. \ref{tab:filament}, it is possible to see that Sector 1 has the lowest temperature; this is consistent with the temperature map of Fig. \ref{figspectral}, which shows that the minimum in temperature is found along the filament. We note that the $\chi^{2}/d.o.f.$ of the fit to Sector 1 is worse than those of the other sectors. It is possible that a fit with a one-temperature model does not represent the multi-phase nature of the gas. In fact, the X-ray emission from filamentary structures in the cores of clusters has typically been analyzed with a multi-temperature approach. \citet{2009MNRAS.393...71S} used three thermal components at fixed temperatures of 0.5, 1.2 and 4 keV to fit the spectrum of the cold blobs in 2A0335+096. \citet{sharma2004} detected gas at $\sim$0.8 keV at the center of Abell~1991 by using the surrounding cluster emission as background (thus removing the ambient thermal gas contribution from the cooler emission); moreover, they tested two-temperature models and thermal$+$cooling flow models to fit the plasma spectrum within $\sim$20 kpc from the center. Similarly, \citet{2011ApJ...732...13G} used a two-temperature model to describe the emission from the filament in Hydra~A. Therefore, we verified whether the inclusion of other components could improve the fit to the filament in ZwCl~235\footnote{We tested the inclusion of these components also for the other sectors, finding that either the $\chi^{2}/d.o.f.$ is worse, or the parameters assume non-physical values.}.
\\ \indent  First, we tested a two-temperature (2T) model (\texttt{tbabs$\ast$(apec+apec)}), linking the abundances of the two thermal component. We found that a second thermal model improves the fit (F-statistics value of 4.8, with a $p$-value of 0.01), and suggests the presence of a thermal component with temperature around 1 keV. \\ \indent  We also tested the inclusion of a cooling flow model (\texttt{tbabs$\ast$(apec+mkcflow)}), linking the higher temperature and abundances of the \texttt{mkcflow} to those of the \texttt{apec}, freezing the lower temperature to 0.08 keV (the minimum allowed by \texttt{XSPEC}) and leaving the mass deposition rate $\dot{M}$ free to vary. We found that for a $\dot{M}\approx11$ M$_{\odot}$ yr$^{-1}$ the $\chi^{2}/d.o.f.$ slightly improves, but not significantly.
\begin{table*}[ht]
	\centering
	\setlength\tabcolsep{5pt}
	\renewcommand{\arraystretch}{1.5}
	\caption{Spectral analysis of the ICM in four sectors (first column) within 3$''$ - 13$''$ from the BCG (see \cref{brightf} and Fig. \ref{figband}, \textit{upper panel} for details).
	}
	\label{tab:filament}
	\begin{tabular}{c|ccc|cccc|cccc}
		\hline
          & \multicolumn{3}{c|}{1T model: \texttt{tbabs$\ast$apec}} & \multicolumn{4}{c|}{2T model: \texttt{tbabs$\ast$(apec+apec)}} & \multicolumn{4}{c}{Cooling model: \texttt{tbabs$\ast$(apec+mkcflow)}} \\
         \cline{2-4} \cline{5-8} \cline{9-12}
		  &       $kT$ & $Z$ & $\chi^{2}/d.o.f.$ &$kT_{1}$ & $kT_{2}$ & $Z$ & $\chi^{2}/d.o.f.$ &$kT$ & $Z$ & $\dot{M}$ & $\chi^{2}/d.o.f.$\\
		  &       [keV] & [Z$_{\odot}$] &  &[keV] & [keV] & [Z$_{\odot}$] &  &[keV] & [Z$_{\odot}$] & [M$_{\odot}$ yr$^{-1}$] & \\
		\hline
		\vspace{0.5mm} S1  & $2.01^{+0.10}_{-0.17}$ & $0.72^{+0.18}_{-0.15}$ & 51/39 & $1.31^{+0.32}_{-0.25}$ & $2.66^{+0.97}_{-0.36}$ & $0.85^{+0.44}_{-0.32}$ & 39/37 & $2.40^{+0.24}_{-0.42}$ & $1.21^{+0.46}_{-0.33}$ & $11.4^{+5.5}_{-3.7}$ & 44/38		\vspace{0.5mm}\\
		\hline
		\vspace{0.5mm} S2  & $2.24^{+0.22}_{-0.23}$ & $0.66^{+0.24}_{-0.21}$ & 31/32 & (...) & (...) & (...) & (...) & (...) & (...) & (...) & (...)		\vspace{0.5mm}\\
		\hline
		\vspace{0.5mm} S3  & $2.72^{+0.27}_{-0.18}$ & $1.08^{+0.44}_{-0.36}$ & 25/23 & (...) & (...) & (...) & (...) & (...) & (...) & (...) &	(...)	\vspace{0.5mm}\\
		\hline
		\vspace{0.5mm} S4  & $3.08^{+0.33}_{-0.32}$ & $0.56^{+0.25}_{-0.21}$ & 28/31 & (...) & (...) & (...) & (...) & (...) & (...) & (...) & (...)		\vspace{0.5mm}\\
		\hline
	\end{tabular}
	\tablefoot{ For the 1T model (\texttt{tbabs$\ast$apec}) we list: (2) temperature; (3) abundance; (4) $\chi$-statistics/degrees of freedom. For the 2T model (\texttt{tbabs$\ast$(apec+apec)}) we list: (5) temperature of the first component; (6) temperature of the second component; (7) abundance (linked between the two components); (8) $\chi$-statistics/degrees of freedom. For the cooling model (\texttt{tbabs$\ast$(apec+mkcflow)}) we list: (9) temperature of the \texttt{apec} component; (10) abundance (linked between \texttt{apec} and \texttt{mkcflow}); (11) mass deposition rate; (12) $\chi$-statistics/degrees of freedom.}
\end{table*}	
\\ \indent  Thus, the spectrum of the filament is best described by two thermal models, one with $kT_{1}\approx1.3$ keV and the other with $kT_{2}\approx2.7$ keV. We notice the similarity of this fit and the one to the X-ray emission of the BCG. In fact, we found a pronounced bump around 0.8-1.0 keV also in the spectrum of the filament, which is distinctive of a plasma with temperature around 1 keV.
\\ \indent To accurately recover the morphology of the cool component within 20 kpc from the center, it would be necessary to map the temperature of the ICM with a 2T model and a spatial resolution of $\sim$2-3$''$, which would require a deeper \textit{Chandra} exposure. Here we apply an alternative, approximate method to trace the distribution of the $\sim$1 keV thermal plasma (see also e.g., \citealt{2013ApJS..206....7M,2017A&A...600A.135B}). We produced background subtracted, exposure corrected images of the ICM in four energy bands: 0.5-0.8 keV (IM1), 0.8-1.2 keV (IM2), 1.2-4 keV (IM3), and 4-7 keV (IM4), which are shown in Fig. \ref{figband} (lower panels). As it is possible to see, the filament is clearly visible only in IM2, while the other images reveal a relatively spherical or amorphous emission around the X-ray peak (green cross). The contours shown in Fig. \ref{figband} start from 3$\sigma$ above the average surface brightness between 3$''$-13$''$ from the center, and nicely trace the region where the bright filament is found. 
\\ \indent  The presence in a sloshing cluster of a cold filament, with spectral properties that resemble those of the BCG's putative thermal corona, represents an interesting opportunity to investigate how the cycle of AGN feeding and feedback is coupled with the dynamics of the central, low entropy gas. We discuss possible explanation for the properties of the filament in \cref{originfil}.

\section{Discussion}
\label{discussione}
In the following, we discuss the results presented in Sect. \ref{result}, considering the information on the ICM properties, the BCG X-ray emission and the activity of the central AGN.  
\subsection{Metal redistribution in the cluster}\label{metaldist}
Our analysis of the ICM metallicity distribution in ZwCl~235 (see \cref{metal}) unveiled 
\begin{figure}[ht]
\includegraphics[width=1\linewidth]{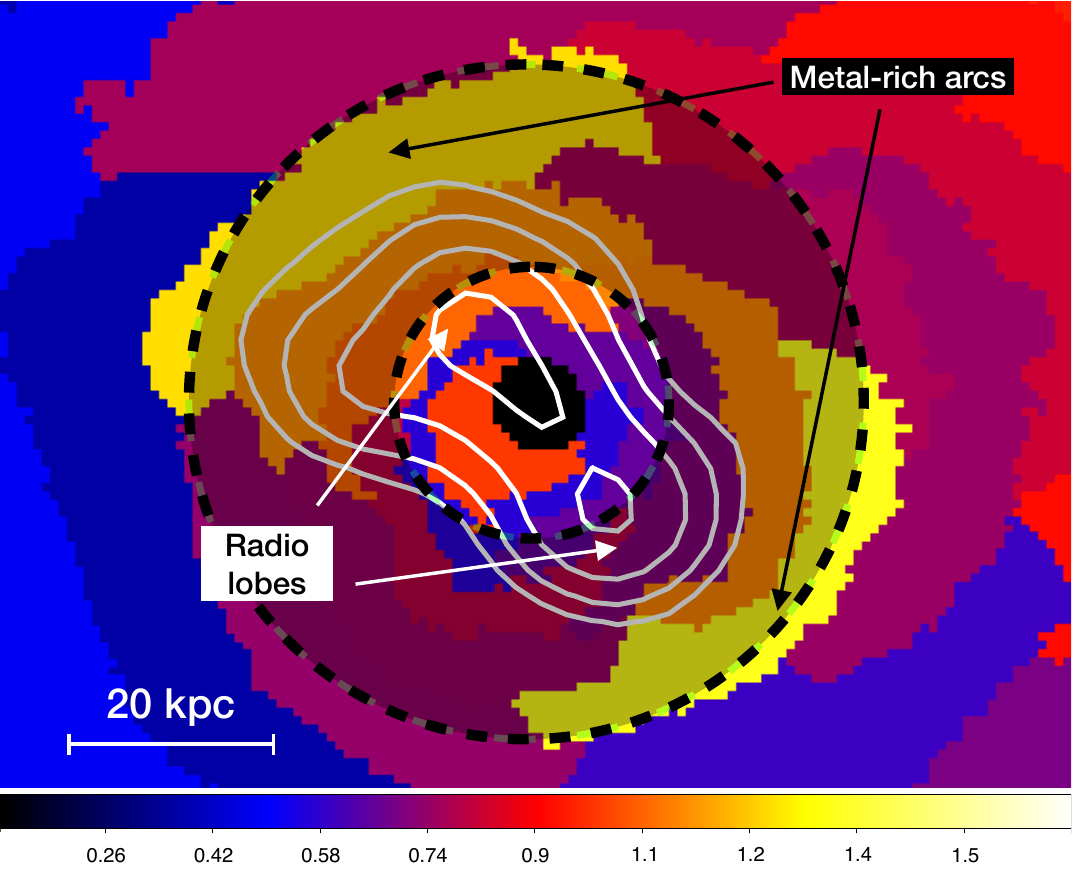}
\caption{Zoom-in of the metallicity map shown in Fig. \ref{figZ}. White contours show the central radio galaxy at 144 MHz (levels are the same of Fig. \ref{figcav_paper}). The translucent black area within the black dashed circles shows the predicted location of the iron radius (see text for details), which matches the position of the metal-rich arcs. The black central region corresponds to the BCG, whose X-ray emission was excluded in creating ICM spectral maps (see \cref{general} and \ref{bcg}).}
         \label{metal_withradio}
\end{figure}
that the metallicity profile shown in Fig.\ref{profile} is relatively flat (with a mean Z$\sim$0.9 Z$_{\odot}$) within $\approx$70 kpc (although it should be noted that the relatively large uncertainties of the profile shown in Fig. \ref{profile} prevent us from drawing firm conclusions on the radial trend of abundances). 
As mentioned in \cref{metal}, this rather flat profile does not meet the expectations for relaxed clusters, in which the enrichment due to the stellar activity of the BCG has determined a prominent peak in iron abundance \citep{2004A&A...416L..21B}. 
We note that sloshing may be an efficient mechanism to broaden the distribution of metals in clusters (see e.g., the analysis of Abell~496 in \citealt{2014A&A...570A.117G}; see also \citealt{2011MNRAS.413.2057R,2014MNRAS.437..730O}). 
Since our analysis indicates that the ICM in ZwCl~235 is sloshing about the center, we may cautiously speculate that an outward transport of metals by the east and west cold fronts could have flattened the metallicity profile within $\sim$70 kpc. 
\\ \indent Besides the putative sloshing-induced asymmetries in iron abundance, the metallicity map and our sector-based spectral analysis of the ICM suggest the existence of two high-metallicity arcs on opposite sides of the core, extending out to 30 kpc from the center. These arcs do not seem to be related to the sloshing mechanism, being misaligned by almost 90$^{\circ}$ with respect to the east and west cold fronts. On the contrary, their position angle of roughly 130$^{\circ}$ is in good agreement with that of the radio-filled X-ray cavities ($\sim$120$^{\circ}$, see \cref{feedback}). Additionally, the $\sim$20-30 kpc distance from the center of the two arcs is of the same order as the radio lobe length as seen at 3 GHz ($\sim$15 kpc) and at 144 MHz ($\sim$20 kpc, see Fig. \ref{figcav_paper} and \ref{metal_withradio}). Therefore, the presence of excess metallicity in the NE-SW direction may be connected to the inflation of the radio lobes.
\\ \indent Observational evidence of metal transport by AGN mechanical outbursts were first found in the Hydra A cluster (see \citealt{2009A&A...493..409S,2009ApJ...707L..69K}); the subsequent study of other systems (see e.g., \citealt{2011ApJ...731L..23K,2015MNRAS.452.4361K}) has allowed to find a relationship between the cavity power $P_{\text{cav}}$ and the altitude of the uplifted gas (the so-called \say{iron radius}), $R_{\text{Fe}}$, in the form (see \citealt{2015MNRAS.452.4361K}):
\begin{equation}
\label{ironradius}
    R_{\text{Fe}}\,\text{[kpc]}= (62\pm26)\times \Bigg(\frac{P_{\text{cav}}}{\text{10$^{44}$ erg s$^{-1}$}}\Bigg)^{(0.45\pm0.06)}
\end{equation}
Using the measured cavity power in ZwCl~235 of 1.2$\times10^{43}$ erg s$^{-1}$ (see \cref{feedback}) yields an iron radius of $R_{\text{Fe}}=24.4\pm9.9$ kpc. As it is possible to see in Fig. \ref{metal_withradio}, the predicted radial range for the iron radius overlaps with the metal-rich arcs. 
Altogether, this evidence may suggest that the metal-rich arcs have been generated by uplift of metals in the cluster atmosphere due to the activity of the central AGN (see for comparison the peculiar distribution of metals in e.g., ZwCl~8276, \citealt{ettori2013}). However, taking into account the various assumptions on the existence of the metal arcs and the cavities, we caution that these results are more speculative. 
\subsection{The origin of the cold gas filament}
\label{originfil}
\begin{figure*}[ht]
\includegraphics[width=1\linewidth]{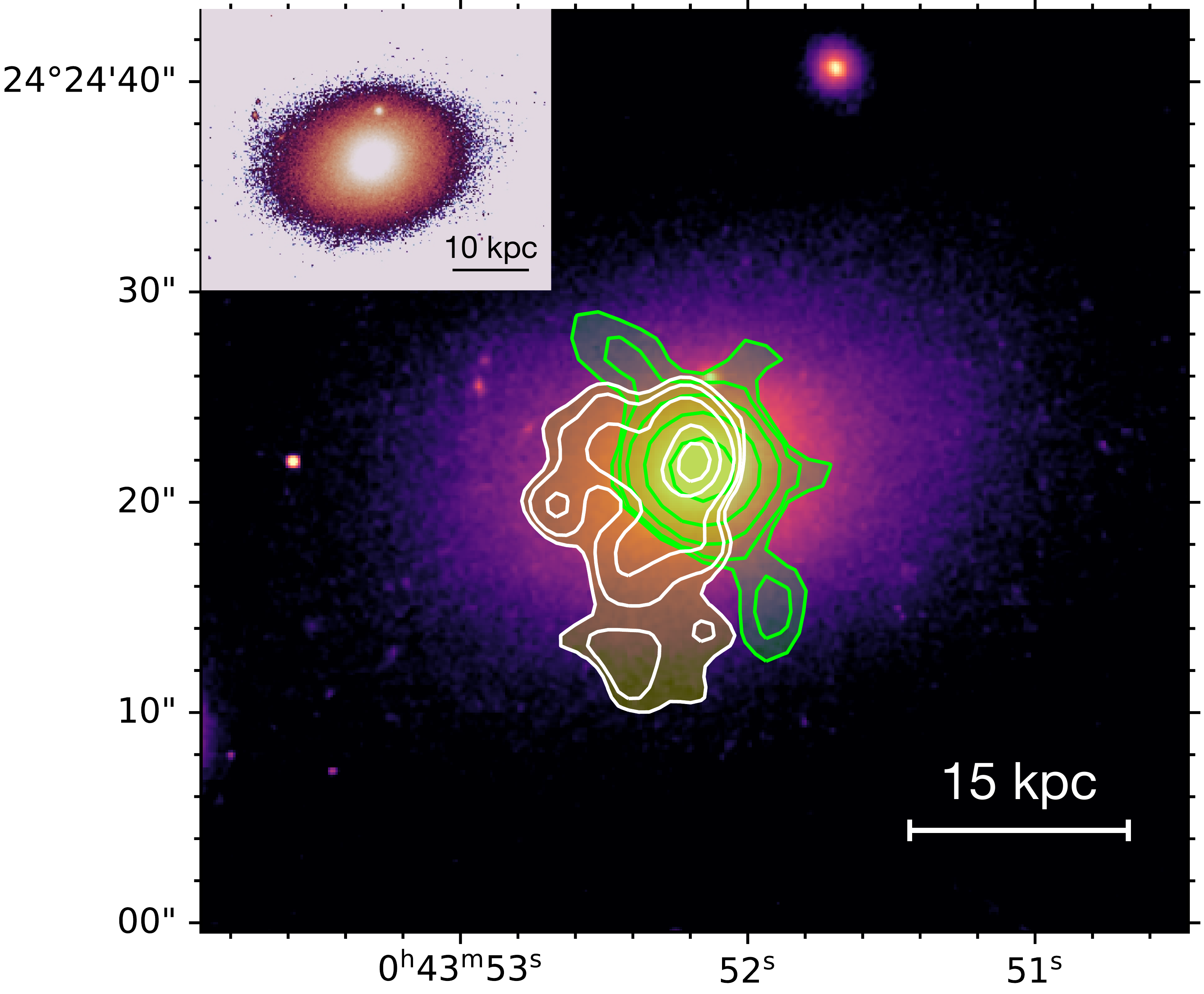}
\caption{Main panel: Multi-wavelength composite image of the center of ZwCl~235. The optical image of the BCG is the \textit{Hubble Space Telescope} image in the broad V filter (WFC F606W). Overlaid in green are filled radio contours at 3 GHz of the central radio galaxy (same as in Fig. \ref{figcav_paper}). Overlaid in yellow are filled X-ray contours between 0.8 - 1.2 keV (same as in Fig. \ref{figband}), showing the bulk of the $\sim$1 keV central thermal plasma (central kernel and bright filament). Upper left panel: zoom-in showing the HST optical image of the BCG.}
         \label{multi}
\end{figure*}

A peculiar feature of the ICM at the center of ZwCl~235 is an arm-like filament, best visible in the soft X-ray band, that extends south of the BCG and is elongated along the southern lobe of the central radio galaxy (see \cref{brightf}). We show in Fig. \ref{multi} a multi-wavelength view of the central regions of ZwCl~235, with the BCG visible in the optical band (image obtained from the HST archive\footnote{\url{https://hla.stsci.edu}}), the radio galaxy contoured in green and the 0.8 - 1 keV X-ray emission of the filament contoured in yellow. 
\\ \indent  We consider three different scenarios that might explain its location, morphology and spectral properties. On the one hand, the filament could represent \textbf{[i]} a cold phase of the ICM, in which cooling has been stimulated due to the activity of the central AGN, that has pushed the gas to an altitude where thermal instabilities might ensue. On the other hand, the gas of the filament could once have been associated with the central gas kernel at 1.4 keV, that may be interpreted as the thermal corona of the BCG (and whose properties are discussed in \cref{bcg}). In this case, the filament could either \textbf{[ii]} represent gas that has been stripped from the BCG due to ram pressure on the corona or \textbf{[iii]} be the result of a small scale \textit{sloshing effect} of the central cool kernel of gas. In the following paragraphs we test and discuss these different hypotheses.

\subsubsection{Stimulated cooling}\label{stimcool}
As reported in the introduction, recent observations of cold filamentary gas from X-ray, molecular, and optical data draped around AGN-inflated lobes and cavities have led to hypothesize that AGN feedback can positively stimulate cooling of the ICM. In such theories, the outbursts lifts cold central gas to an altitude where cooling becomes unstable, leading to condensation of small clouds of gas that rain back onto the BCG. There is not a general consensus on the criterion for thermal instabilities to ensue. One the one hand, it has been suggested that unstable cooling occurs within a region where the ratio between cooling time and free fall time reaches a value of $\sim$10-20 (e.g., \citealt{2015ApJ...799L...1V}), or less than $70$ \citep{2015MNRAS.448.1979V}. The free fall time is defined as:
\begin{equation}
    \label{tfreefall}
        t_{ff}(r) = \sqrt{\frac{2r}{g(r)}}  
\end{equation}
\noindent where $g(r)$ is the gravitational acceleration at distance $r$.
\\ \indent On the other hand, \citet{2018ApJ...854..167G} proposed that cooling instabilities are governed by the eddy turnover time $t_{\text{eddy}}$, that is the time at which a turbulent vortex gyrates and produces density fluctuations, and can be defined as:
\begin{equation}
    \label{teddy}
    t_{\text{eddy}} = 2\pi\,\frac{r^{2/3}\,L^{1/3}}{\sigma_{v}}
\end{equation}
\noindent where $L$ is the injection scale of turbulence (usually assumed to be described by the diameter of the X-ray cavity), and $\sigma_{v}$ is the velocity dispersion of the gas. According to \citet{2018ApJ...854..167G}, the gas should become multi-phase when the ratio between $t_{\text{cool}}$ and $t_{\text{eddy}}$ is close to unity or lower than one. \\ \indent Additionally, it has been proposed that cooling is primarily driven by the local values of entropy and cooling time, specifically $K\leq$30 keV cm$^{2}$ and $t_{\text{cool}}\leq$1 Gyr \citep{2008ApJ...683L.107C}. Below these thresholds, which are not related to local dynamical times but only to thermodynamical properties, the presence of multi-phase gas and rapid cooling is expected (see e.g., \citealt{2016ApJ...830...79M,2017ApJ...851...66H}).  
\\ \indent We show in Fig. \ref{multi} how the positive feedback scenario might apply to ZwCl~235. As it is possible to see, the bright X-ray filament is tangential to southern radio lobe. As discussed in \cref{general}, the azimuthal minimum in temperature and entropy is found between 8$''$ - 12$''$ from the center, which corresponds to the projected length of the filament. Additionally, the cooling time map shown in Fig. \ref{figspectral} confirms that $t_{\text{cool}}$ drops to the center and has its minimum at the position of the filament. We note that the filament appears to be composed of bright clumps (see \textit{upper panel} in Fig. \ref{figband}), that might be cold and dense cooling sites. The best fit to the filament X-ray emission (see \cref{brightf}) confirmed that the local ICM is cold, being at a temperature of $\sim$1.3 keV. With a volume of the cold gas component of 3.2$\times10^{4}$ kpc$^{3}$ and an electron density of $\sim$0.06 cm$^{-3}$ (see Appendix \ref{appendix:volume}), the entropy and cooling time are found to be K$\approx9.7$ keV cm$^{2}$ and $t_{\text{cool}}\approx480$ Myr.
We note that these values meet the thresholds of K$\leq$30 keV cm$^{2}$ and $t_{\text{cool}}\leq$1 Gyr for the onset of cooling. \\ \indent In order to test the $t_{\text{cool}}/t_{ff}$ criterion, we computed the free fall time using Eq. \ref{tfreefall} and the mass profile of ZwCl~235 (see \cref{feedback}) at the average distance of the filament from the center, i.e. roughly 12 kpc. With a free fall time of $t_{ff}\approxeq29$ Myr, we estimate that $t_{\text{cool}}/t_{ff}\sim17$. \\ \indent As a last check, we computed the eddy timescale (Eq. \ref{teddy}) by assuming that the cavity diameter (6.5 kpc) is representative of the injection scale $L$. We note that the cavity diameter represents a lower limit to the injection scale $L$, as turbulence could be injected on the whole cold gas region ($\sim$20 kpc). Regarding the velocity dispersion, \citet{2018ApJ...854..167G} showed that the different gas phases (hot, warm, neutral and molecular) are linked in terms of $\sigma_{v}$ in the inner $\sim$50 kpc of clusters and groups, with velocity dispersions ranging between 100-200 km s$^{-1}$. Observationally, this indicates that it is possible to use the condensed gas $\sigma_{v}$ as a tracer of the ICM velocity dispersion. By analyzing the kinematics of the molecular gas in a sample of 15 clusters, \citet{2019A&A...631A..22O} found $\sigma_{v}$ in the range 104 - 196 km s$^{-1}$. Since ZwCl~235 does not have measurements of the multi-phase gas kinematics, we assume $\sigma_{v}=200$ km s$^{-1}$. We note that the assumptions of $L=6.5$ kpc and $\sigma_{v}=200$ km s$^{-1}$ are driven by the aim of being conservative, knowing that larger scales and/or a lower velocity dispersion would determine a longer eddy timescale and a smaller $t_{\text{cool}}/t_{\text{eddy}}$ ratio. At a distance $r$ of 12 kpc from the center we measure $t_{\text{eddy}}\sim302$ Myr, and thus $t_{\text{cool}}/t_{\text{eddy}} \sim 480/302 \sim 1.6$, which is close to unity and supports the existence of multi-phase gas. Nonetheless, due to the several assumptions, we consider our measurements of $t_{\text{cool}}$, $K$ and $t_{\text{cool}}/t_{ff}$ as our most robust indicators of stimulated cooling.
\\ \indent Overall, we find that ZwCl~235 meets the thermodynamical thresholds and the dynamical criteria. The end product of ICM cooling would be the co-existence of different gas phases over a wide range of temperatures. Probing the presence of such components typically requires either spatially resolved optical line observations (showing extended H$\alpha$ nebular emission, see e.g., \citealt{2019A&A...631A..22O}) or radio observations targeting molecular gas (e.g. ALMA data of CO lines, see e.g., \citealt{2019MNRAS.490.3025R}). Such probes are currently unavailable for ZwCl~235, although hints for the existence of multi-phase gas in its core can be found in the literature. As reported in the introduction, \citet{1999MNRAS.306..857C} measured a relatively high H$\alpha$ luminosity of L$_{H\alpha}=(4.1\pm0.6)\times10^{40}$ erg s$^{-1}$, which could be indicative of extended warm gas nebulae at $\sim10^{4}$ K (e.g., \citealt{2016MNRAS.460.1758H}). However, spatially resolved spectroscopy is required to investigate the morphology of the warm gas. Regarding molecular gas, the high resolution optical image in Fig. \ref{multi} (upper left panel) does not show evidence for dust lanes in absorption or star forming knots, which indicates that the molecular gas content may not be extreme (for molecular rich clusters see e.g., Abell~1795, \citealt{russell2017}, or 2A0335+096, \citealt{2016ApJ...832..148V}). In this respect, \citet{2003A&A...412..657S} found hints of a CO(1-0) line in the core of ZwCl~235 using an IRAM 30 m observation, and placed an upper limit on the molecular mass of M$_{\text{mol}}\leq2\times10^{9}$ M$_{\odot}$. It is interesting to note that the test of fitting the filament's spectrum with a cooling model (see Tab. \ref{tab:filament}) returned a non-null value for the mass deposition rate of $\dot{M}\approx11$ M$_{\odot}$ yr$^{-1}$. Interpreting this as an upper limit to the star formation rate, the M$_{\text{mol}}\leq2\times10^{9}$ M$_{\odot}$ estimated by \citet{2003A&A...412..657S} would be depleted in $\gtrapprox 200$ Myr, which is typical of other BCGs surrounded by multi-phase gas (see e.g., \citealt{2016A&A...595A.123M,2016MNRAS.458.3134R,2020A&A...640A..65C,2021ApJ...910...53V}). 
\\ \indent  Overall, we expect more sensitive optical and radio observations to be able to detect and resolve extended molecular gas, and test the hypothesis that the X-ray filament is undergoing condensation into cold gas clouds. 
\subsubsection{Ram pressure stripping}\label{bcgram}
Through the process of ram pressure stripping, the ICM can be efficient in removing the hot coronal component of ellipticals, generating X-ray emitting tails behind the galaxy. While the process is most evident in the case of star-forming galaxies (the so-called \say{jellyfish} galaxies, see e.g., \citealt{2017ApJ...844...48P,2021arXiv210913614B}), even elliptical galaxies moving in clusters can be subject to extreme stripping of their halos (for a recent example see \citealt{2022MNRAS.511.3159M}). In \cref{brightf} we noticed the similarity between the spectral properties of the filament ($kT\sim1.3$ keV, Z$\sim0.9$ Z$_{\odot}$) and those of the BCG thermal halo ($kT\sim1.4$ keV, Z$\sim0.8$ Z$_{\odot}$). As it is possible to see from Fig. \ref{figband} and \ref{multi}, the X-ray emission between 0.8 - 1.2 keV shows that the filament is connected to the central source and extends in the south direction. Therefore, it is possible that the gas in the filament originates from the BCG gas halo. In particular, the X-ray tail might have been created in the past after ram pressure stripping of the kernel's outer layers ($\gtrsim$5 kpc). 
\\ \indent It should be noted that the stripping is efficient when galaxies are moving at a relatively high speed w.r.t. the cluster center; in the case of BCGs, their location at the cluster center indicates that relative motions, if present, are not significant. This applies to ZwCl~235, in which the BCG is coincident with the cluster center. An interesting exception is constituted by the BCG of MKW08 \citep{2019A&A...629A..82T}, that shows a 40 kpc-long X-ray tail; in this case, however, the host cluster is gravitationally interacting with the galaxy group MKW07, suggesting that residual motions might be present. On the other hand, projection effects might be hiding a significant offset between the BCG of ZwCl~235 and the cluster center, which would imply the presence of relative motions aligned with the line of sight. These motions would then generate the conditions for efficient ram pressure stripping to occur. Nevertheless, if this was the case we would not expect to observe the long X-ray tail ($\sim$20 kpc), which would be equally projected along the line of sight. Therefore, we argue that a line of sight motion of the BCG is unlikely. 
\\ \indent From the observational point of view, if the tail was stripped from the BCG halo we expect millimeter observations to reveal that the molecular gas is still coincident with the central emission. While this information is currently unavailable for ZwCl~235, we can make comparison to the similar clusters reported in \cref{subsec:z235}. In particular, due to the typically higher densities of molecular gas ($\sim$10$^{2}$ cm$^{-3}$, e.g., \citealt{2019MNRAS.490.3025R,2021A&A...649A..23C}) w.r.t. those of the hot gas ($\sim$0.1 cm$^{-3}$), the cold phase is more difficult to strip and would require relative velocities of $\geq$1000 km s$^{-1}$ (see e.g. \citealt{2019ApJ...870...57V}). Considering the central position of the BCG, it is very unlikely that it is moving at such supersonic speed (for ZwCl~235 it would imply a Mach number of $\sim$2). In this respect, similar conclusions have been drawn for other X-ray bright tails of gas connected to cluster-central BCGs; restricting the comparison to the systems in Tab. \ref{tab:objects}, in Abell~1795 \citep{russell2017}, 2A0335+096 \citep{2016ApJ...832..148V} and Abell~1991 \citep{hamer2012}, the co-spatiality of the cool ICM and of the molecular gas disfavors the ram pressure stripping scenario, and supports instead a local multi-phase condensation of the hot gas.
\\ \indent Observational evidence for the coronal origin of tails can also include higher metallicity along the tail w.r.t. the ambient gas (see e.g., \citealt{2019A&A...629A..82T}). As detailed in \cref{metal}, the distribution of metals in ZwCl~235 has possibly been perturbed by sloshing and by AGN uplift. This generates significant difficulties in ascribing any metal asymmetry within $\sim$15 kpc from the center to other mechanisms such as ram pressure stripping. While from the metallicity map shown in Fig. \ref{metal_withradio} there are hints of higher metallicity to the south-east side of the BCG, we are unable to further investigate this point due to the short exposure and the overall complex abundance distribution. 
\subsubsection{Sloshing of the BCG corona}\label{bcgslosh}
Following gravitational perturbations that offset the densest ICM from the bottom of the potential well of the cluster, the sloshing motion generates spiral patterns in the cold gas distribution. This scenario applies to the ICM in ZwCl~235: the three cold fronts reported in \cref{fronts} trace the spiral geometry of the gas on scales from 30 kpc (the east front) to 100 kpc (south-east front). Here we consider the effect that this perturbation may have had on the cooler, denser plasma at the cluster center. In particular, we investigate the idea that besides generating the three cold fronts in the ICM, sloshing may have shaped the outer layers of the BCG's thermal corona (or the most central ICM phase), leading to the formation of a spiral-tail of gas with kT$\sim$1 keV.
\\ \indent As already mentioned, the spectral properties of the bright X-ray filament resemble those of the central thermal plasma at 1.4 keV. Furthermore, the cold filament seen in the 0.5 - 2 keV image (Fig. \ref{figchandra}), the temperature map (Fig. \ref{figspectral}) and the 0.8 - 1.2 keV contours shown in Fig. \ref{multi} is characterized by a hooked morphology: starting from the BCG it heads south-east then bends to the south. 
The arched structures produced by the outward sloshing motion are expected to be over-dense and colder w.r.t. the surrounding gas, while maintaining pressure equilibrium \citep{2007ARA&A..45..117M}. We showed in \cref{brightf} that the filament is cold and dense compared to the ambient medium, and we estimate a pressure ratio between the filament and the surrounding ICM (from the profiles of Fig. \ref{profile}) of $\sim$1.1, which supports the pressure equilibrium.
Therefore, the filament could represent a small scale ($\sim$20 kpc in projected length) sloshing tail of gas, that was produced by the same event that has set the oscillating movement of the ICM on tens of kpc scales. 
\\ \indent It is possible to provide an estimate of the time taken by the $\sim$1 keV gas to slosh to its observed distance from the center. Based on literature examples, there are three methods that can be used to constrain the sloshing timescale:
\begin{itemize}
    \item \textit{Free fall time}: if there is no outward pressure to counteract the infall of gas, then the time required for a clump of gas at a distance $r$ to return to the center is given by the free fall time $t_{ff}$. As reported in \cref{stimcool}, at a distance of 12 kpc the free fall time is $t_{ff}\approx29$ Myr, which we interpret as a lower limit on the true sloshing timescale (see e.g., \citealt{2021MNRAS.503.4627U}).  

    \item \textit{Brunt-V{\"a}is{\"a}l{\"a} time}: recent studies of sloshing clusters have approximated the motion of the gas around the center as an oscillating flow in a stable environment (e.g., \citealt{2017ApJ...851...69S,2020MNRAS.496.1471K,2021MNRAS.503.4627U,pasini2021}), governed by the Brunt-V{\"a}is{\"a}l{\"a} timescale \citep{1990ApJ...357..353B}:       
        \begin{equation}
            t_{\text{BV}}(r) = 2\pi\,\Bigg( \frac{3g(r)}{5r}\delta_{K} \Bigg)^{-1/2} 
        \label{omegabv}
        \end{equation}
        \noindent where $\delta_{k}$ is the slope of the entropy profile. Since we do not have the statistics to build an entropy profile for the coronal gas, we assume the value of $\delta_{K} = 2/3$ (see the universal entropy profile discussed in \citealt{2018ApJ...862...39B}). For the filament in ZwCl~235 we thus measure $t_{\text{BV}}\approx190$ Myr. As expected from Eq. \ref{omegabv}, the Brunt-V{\"a}is{\"a}l{\"a} timescale is a factor of roughly 6 longer than the free fall time, and can be considered as an upper limit on the sloshing time. 
        
    \item \textit{Subsonic timescale}: the oscillation of gas in sloshing clusters has generally been described as subsonic \citep{2016JPlPh..82c5301Z}. Assuming that the filament gas has traveled at $\sim$0.5$c_{s}$ (see e.g. \citealt{2014MNRAS.437..730O,2022arXiv220104591B}), its speed would be $\sim$370 km s$^{-1}$ (half the value of the sound speed at the cavity distance from the center, see \cref{feedback}). Thus, it would have taken $\approx$35 Myr to reach a distance of 12 kpc from the center, which is close to the free fall time. 
\end{itemize}
Using the above three methods, we constrain the sloshing timescale for the filament to be between 30 - 190 Myr. While this range is relatively broad, it is significantly longer than the age of the AGN outburst ($\sim$20 Myr), which implies that if the central $\sim$1.4 keV thermal kernel has been subject to sloshing, this happened before the formation of the two cavities. 
\\ \indent  The sequentiality of the two events opens the possibility that a combination of sloshing and stimulated cooling might explain the filament origin. In particular, a perturbation of the gas might have offset the central thermal gas at $\sim$1.4 keV to the south-east of the BCG between 30 - 190 Myrs ago. Considering the volume and electron density of the filament (see \cref{stimcool}), the total sloshed mass would be $\sim2\times10^{9}$ M$_{\odot}$, i.e. twice the current mass of the corona. Then approximately 17 Myr ago the AGN started its activity and inflated two bipolar lobes in the NE - SW direction. As the majority of $\sim$1.4 keV gas had already been displaced from the center, the uplift of gas behind the X-ray cavities occurred only where gas to be uplifted was effectively available, i.e. around the south radio lobe. This argument would explain why the low entropy gas is not found around the north radio lobe. Such a scenario holds for example for 2A0335+096, where the presence of a single filament in the direction of one of the cavities suggests that sloshing may have set a preferred direction for cooling of the ICM \citep{2016ApJ...832..148V}. In Abell~1795, the 40-kpc long cooling wake is likely the result of cooling stimulated by the relative motion of the BCG and the ICM \citep{2005MNRAS.361...17C,russell2017}; additionally, in Abell~2052 (that shows signs of sloshing, \citealt{blanton2011}) there is a strong asymmetry in the warm gas distribution, with nebular filaments being draped only around the northern inner cavity \citep{2018A&A...612A..19B}. On the contrary, in the case of e.g. the Phoenix cluster, the warm gas is draped around both the central X-ray cavities (see \citealt{2015ApJ...811..111M}). The inclusion of sloshing can also alleviate the tension between the maximum mass of the gas displaced by the cavity and the actual mass of the 1 keV gas: following Archimedes' principle, a bubble can lift no more mass than the mass it has displaced. From the volume of the cavity reported in Tab. \ref{tab:cavities} and the density of the ICM at 12 kpc from the center ($\sim$0.04 cm$^{-3}$), the maximum amount of displaceable gas would be approximately $3\times10^{8}$ M$_{\odot}$. On the contrary, the mass of the filament is six times higher. Since sloshing can emulate the effect of uplift and provide an alternative way of triggering unstable condensation (see \citealt{2019ApJ...870...57V}), it is plausible that the cavity expansion is responsible for only a small portion of the uplifted gas. Overall, this scenario provides a consistent explanation for the spectral and morphological properties of the cold filament and its connection to the AGN outburst. 
\subsection{Cooling and feedback from the cluster to the BCG scales}
\label{coolingfeedback}
The X-ray and radio analysis of ZwCl~235 revealed a complex system, where the dynamics of the ICM is strongly coupled with the activity of the central AGN. Besides the sloshing cold fronts, we discovered two radio-filled putative X-ray cavities at $\sim$15 kpc from the BCG, which is surrounded by a thermal halo of gas at $\sim$1.4 keV (cooler than the ambient ICM) and with a radius of $\sim$5 kpc. Starting from this cool kernel, a bright filament of gas (at $\sim$1 keV) extends to the south, parallel to the southern radio lobe (see Fig. \ref{multi}). In this section we consider how these different features can be interpreted in the context of AGN feeding and feedback mechanisms. Given the different physical scales involved, we refer to the distinction proposed by \citet{2020NatAs...4...10G} between the \textit{macro} (10s - 100s kpc) and \textit{meso} (pc - kpc) scales of AGN feedback.
\\ \indent  The radio observations of the central radio galaxy, detailed in \cref{radiogalaxy}, reveal that the AGN that inflated the 15 kpc-long radio lobes seen at 3 GHz is currently active: the VLBA image at 5 GHz shows a bright core with bipolar jets emanating in the NE - SW direction. Our question is \textit{which sources of fuel are available for the central SMBH?} 
\\ \indent  Classical pictures of AGN feeding in cool core galaxy clusters predict that cooling of the ICM can provide a resource of cold gas to be accreted by the central engine. In turn, the SMBH launches jets that heat the surrounding medium via cavity inflation and shocks, preventing over-cooling of the ICM and establishing a finely-tuned cycle (see Sect. \ref{intro}). The luminosity of the cooling region, $L_{\text{cool}}$, can be used as a proxy for the rate at which cold gas flows to the center and is accreted, $\dot{M}_{\text{cool}}$, via the expression:
\begin{equation}
    \label{classicflow}
    \dot{M}_{\text{cool}} = \frac{2}{5}\frac{\mu m_{\text{p}}}{kT} L_{\text{cool}}
\end{equation}
From the properties of the cooling region in ZwCl~235 reported in \cref{general}, we estimate a classical cooling rate of $\dot{M}_{\text{cool}}=147\pm8.3$ M$_{\odot}$ yr$^{-1}$. This is the expected rate of gas cooling in the absence of heating sources. Given the scaling between the cooling luminosity and the H$\alpha$ luminosity for systems in Tab. \ref{tab:objects} mentioned in \cref{general}, and by the definition of $\dot{M}_{\text{cool}}$, it is unsurprising to find that systems with lower L$_{H\alpha}$ have mass deposition rates estimated with Eq. \ref{classicflow} of a few tens of M$_{\odot}$ yr$^{-1}$ (as Abell~2495 and Abell~1668, \citealt{2019ApJ...885..111P,pasini2021}), while more H$\alpha$ luminous systems have $\dot{M}_{\text{cool}}\approx10^{3}$ M$_{\odot}$ yr$^{-1}$ (as Abell 1835, \citealt{mcnamara2006}, or Abell 2204, \citealt{2009MNRAS.393...71S} and references therein). \\ \indent In the presence of central activity, the cavity power is usually assumed to represent a lower limit estimate of the total mechanical power of the AGN. The predictions for the feedback cycle expect the cavity power to be approximately one order of magnitude within the cooling luminosity (or $pV/t_{\text{age}}\lessapprox L_{\text{cool}}\lessapprox16pV/t_{\text{age}}$, see e.g., \citealt{2004ApJ...607..800B,2006ApJ...652..216R,2011ApJ...735...11O}). This is indeed the case for ZwCl~235, where we estimated $P_{\text{cav}}=4pV/t_{\text{age}}=1.2\times10^{43}$ erg s$^{-1}$ and $L_{\text{cool}}\approxeq10^{44}$ erg s$^{-1}$, and for the clusters in Tab. \ref{tab:objects} with measurements of both $P_{\text{cav}}$ and $L_{\text{cool}}$ (namely, Abell~1668, \citealt{pasini2021}; Abell~2495, \citealt{2019ApJ...885..111P}; Abell~2204, \citealt{2009MNRAS.393...71S}; Abell~2052, Abell~1835, 2A0335+096, Abell~478, Abell~2199, Abell~1795, e.g., \citealt{2006ApJ...652..216R}). Therefore, by investigating the feedback cycle on macro scales (between approximately 20 - 100 kpc), we confirm the coupling of the central radio activity with the thermodynamic properties of the ambient gas, as extensively tested in literature studies (e.g., \citealt{2004A&A...416L..21B,2006ApJ...652..216R,2015ApJ...805...35H}). 
\linebreak
\\ \indent The above considerations are based on the azimuthally-averaged profiles shown in Fig. \ref{profile}. However, as discussed in \cref{brightf}, the bulk of the coolest gas in ZwCl~235 is not uniformly distributed around the core, as expected from a spherically-symmetric inward flow of cold gas. Instead, $\sim$1 keV gas of low entropy ($\leq$30 KeV cm$^{2}$) and short cooling time ($\leq$0.5 Gyr) is preferentially found in the south direction, along a bright filament (length of $\approx$20 kpc) that is tangential to the southern radio lobe. As discussed in \cref{stimcool}, the alignment of the radio galaxy with the lowest entropy medium is likely explained in the context AGN-stimulated condensation of gas in low entropy clouds, with a possible contribution from sloshing (see \cref{bcgslosh}). This places ZwCl~235 in good agreement with some well-studied systems in Tab. \ref{tab:objects} (e.g., 2A0335+096 or Abell~1795) where sloshing may have set a preferential direction for AGN-stimulated cooling to occur. According to recent simulations of AGN feeding and feedback (see \citealt{2018ApJ...854..167G}), the cold gas kernels along filaments would ultimately provide an optimal fuel for the central SMBH. Hence, even on smaller scales (between 5 - 20 kpc) the fueling of the central engine and the growth of the radio galaxy in ZwCl~235 are found to be coupled with the cooling conditions of the ICM.
\linebreak
\\ \indent  The observations of AGN activity on meso scales (tens of pc) with the VLBA (see Fig. \ref{figcav_paper}) show that the radio galaxy is active and is currently driving jets through its surroundings. While the study of the macro scales (20 - 100 kpc) supports the idea that feeding of the SMBH comes from cooling of the ICM, the BCG thermal halo (see Sect \ref{bcg}) is a further piece of the puzzle. 
This dense ($n_{e}\sim0.1$ cm$^{-3}$) and extended ($\sim$5 kpc in radius) gas kernel at 1.4 keV is the gas phase in ZwCl~235 with the lowest entropy (K$\sim$7 keV cm$^{2}$) and the shortest cooling time ($t_{\text{cool}}\sim240$ Myr). In the absence of the outer environment, the thermal halo (that we interpreted as the BCG's corona) would thus be the closest source of feeding for the central SMBH. However, the ICM outside $r\sim5$ kpc in ZwCl~235 seems to be regulating the amount of fuel for the AGN and the resulting energy output. Therefore, it is necessary to understand how the central cool halo of the BCG is coupled with the ICM/AGN feedback cycle. Besides AGN feedback, stellar mass loss and supernovae explosions are likely to contribute to heating of the thermal halo of cluster-central elliptical galaxies (e.g., \citealt{2009ApJ...704.1586S}). The energy injection required from the AGN should be strong enough to reduce cooling without completely destroying the corona. We note that the mechanical AGN power in ZwCl~235 is $\sim$20 times larger than the X-ray bolometric luminosity of the corona (see \cref{feedback} and \cref{bcg}). In order for the corona to survive such outburst and be observable, the majority of the AGN mechanical power has likely been deposited outside the BCG ($\geq5$ kpc; see also the discussion in \citealt{2009ApJ...704.1586S}). 
\\ \indent We stress that ZwCl~235 differs from the typical systems in the sample of \citet{2009ApJ...704.1586S}, which comprises coronae of weak cool core clusters ($t_{\text{cool}}\geq$1 Gyr) and of non-cool core clusters; in these objects the $\sim$1 keV gas halo of the BCG naturally explains how the central AGN is being fueled even though the outer environment is not cooling (see also \citealt{2019A&A...629A..82T}). A fitting example is given by Abell~2634 (one of the lowest H$\alpha$-luminosity systems listed in Tab. \ref{tab:objects}), where the BCG's corona at 1.0 keV is surrounded by a $\sim$4 keV ICM \citep{schindler1997,2007ApJ...657..197S}. In Abell~2634 the corona is the only thermal gas phase with a density, entropy and cooling time that support the existence of cold gas at the cluster core; using the bolometric X-ray luminosity of the 1 keV plasma, \citet{schindler1997} measured a classical mass deposition rate of $\sim$1 M$_{\odot}$ yr$^{-1}$. On the contrary, ZwCl~235 is a cool core galaxy cluster ($t_{\text{cool}}\leq$1 Gyr within the central 25 kpc) that additionally hosts a dense X-ray corona at its center. By applying the standard cooling flow model (Eq. \ref{classicflow}) to the properties of the corona, that has a bolometric luminosity $L_{\text{cool}}^{BCG}=(6.2\pm1.3)\times10^{41}$ erg s$^{-1}$ and a temperature of $kT\sim1.4$ keV, we estimate a $\dot{M}_{\text{cool}} = 1.6\pm0.9$ M$_{\odot}$ yr$^{-1}$ (similar to that of the corona in Abell~2634), i.e. the cooling rate within 5 kpc is about 1\% of the ICM cooling rate between 5 kpc and $\sim$79 kpc (the cooling radius of ZwCl~235, see \cref{general}). Thus, the thermal corona is not an \say{isolated fuel reservoir}: both the BCG thermal halo and the central ICM have temperature, entropy and cooling time that may trigger the condensation of 1-2 keV gas into colder gas clouds. Concerning fueling of the central SMBH, the two components may as well be considered as a single multi-temperature gas, with the coldest phase of the ICM being indistinguishable from the corona outer layers. 
\\ \indent Overall, we conclude that both the ICM and the thermal halo of the BCG are sources of multi-phase, condensed gas in the inner $\sim$20 kpc of the cluster, with the ICM likely building the larger reservoir.


\section{Conclusions}
\label{conclusione}
In this article we investigated the AGN feeding and feedback cycle in the galaxy cluster ZwCl~235 using available X-ray (\textit{Chandra}) and radio (LOFAR, VLA, VLBA) observations of the central radio galaxy. Here we summarize our results.
   \begin{enumerate}
      \item  The morphological and spectral analysis of the \textit{Chandra} observation unveiled that ZwCl~235 is a cool core galaxy cluster ($t_{\text{cool}}\leq$1 Gyr within 25 kpc from the center, cooling radius $r_{\text{cool}}=78.5\pm1.4$ kpc, cooling luminosity $L_{\text{cool}}\approxeq10^{44}$ erg s$^{-1}$) experiencing sloshing of its ICM, that resulted in the formation of three cold fronts wrapped around the center. The central radio galaxy of ZwCl~235 is currently active: a VLBA observation at 5 GHz revealed a bright core and a pair of jets with a projected total length of 20 pc. The jets are aligned with the radio lobes seen at 144 MHz and 3 GHz by the LOFAR and VLASS surveys, respectively, that extend in the NE - SW direction for approximately 15-20 kpc. We found a pair of X-ray cavities in the ICM matching the position of the radio lobes of the central AGN, whose combined mechanical power is approximately 1.2$\times10^{43}$ erg s$^{-1}$. 
      \item The \textit{Chandra} spectral analysis of the inner 3$''(4.7$ kpc) revealed that the X-ray emission originates from a kernel of thermal plasma at $\sim$1.4 keV coincident with the BCG, with the contribution of non-thermal emission being negligible. The plasma has a density of $\sim$0.1 cm$^{-3}$ and an entropy of 7 keV cm$^{2}$, and is in pressure equilibrium with the surrounding hotter ($\sim$2.7 keV) environment. We interpret this component as the BCG's thermal halo/corona.
      \item Our analysis of the metallicity in ZwCl~235 hints at a relatively flat abundance radial profile (with a mean Z$\sim$0.9 Z$_{\odot}$ within $\sim$70 kpc from the center) and two enhanced metallicity arc-like regions beyond the AGN radio lobes (extending out to $\sim$30 kpc from the center). We argue that this distribution of metals may reflect the ICM dynamical perturbations triggered by sloshing and AGN activity. Sloshing may be responsible for an outward transport of metals, traced by the abundance gradients associated with the east and west cold fronts. The central AGN, while inflating the X-ray cavities, possibly generated a pair of metal-rich arcs at 30 kpc from the center in the NE - SW direction by pushing central enriched material to higher altitudes. 
      \item We discovered that the bulk of the cool ICM is found along an X-ray filamentary feature that starting from the center extends to the south for approximately 20 kpc. A double thermal model best describes the emission of the filament, with the cooler phase having $kT\sim1.3$ keV, $n_{e}\sim0.06$ cm$^{-3}$, and $t_{\text{cool}}\sim480$ Myr. We tested three scenarios for the origin of the filament, namely [i] stimulated cooling by uplift of cool gas, [ii] ram pressure stripping of the BCG coronal gas, or [iii] sloshing of the BCG thermal halo. By considering the properties of the filament, we concluded that a combination of sloshing and stimulated cooling might provide the best explanation for its origin. 
      \item By comparing the cooling properties of the X-ray thermal gas (ICM + BCG thermal halo), we discussed how the cycle of AGN feeding and feedback acts on different scales. Between 20 - 100 kpc from the center (macro scales), we confirm standard predictions of the feedback cycle, with the luminosity of the cooling region being related to the mechanical energy released by the AGN outburst. Between 5 - 20 kpc from the center (meso-macro scales) the thermodynamic state of the ICM is found to be strongly coupled with radio activity, as demonstrated by the cold filament trailing the southern X-ray cavity. On scales of $\leq$5 kpc (meso scales), we speculate that the thermal halo of the BCG is contributing to the inflow and cooling of hot gas to lower temperature gas phases, with the ICM building the larger reservoir of fuel for the SMBH.
      Therefore, we propose that the AGN is possibly linked with multiple sources of material to be accreted.
   \end{enumerate}


\begin{acknowledgements}
    We are grateful to the referee Alastair Edge for the valuable comments that improved the paper and his helpful suggestions. We acknowledge financial contribution from the agreement ASI- INAF n.2017-14-H.0 (PI: A. Moretti).
    This research has made use of data obtained from the Chandra Data Archive and the Chandra Source Catalog, and software provided by the Chandra X-ray Center (CXC) in the application packages CIAO and Sherpa.
    The National Radio Astronomy Observatory is a facility of the National Science Foundation operated under cooperative agreement by Associated Universities, Inc.
    LOFAR data products were provided by the LOFAR Surveys Key Science project (LSKSP; https://lofar-surveys.org/) and were derived from observations with the International LOFAR Telescope (ILT). LOFAR (van Haarlem et al. 2013) is the Low Frequency Array designed and constructed by ASTRON. It has observing, data processing, and data storage facilities in several countries, which are owned by various parties (each with their own funding sources), and which are collectively operated by the ILT foundation under a joint scientific policy. The efforts of the LSKSP have benefited from funding from the European Research Council, NOVA, NWO, CNRS-INSU, the SURF Co-operative, the UK Science and Technology Funding Council and the Jülich Supercomputing Centre.
    This research is based on observations made with the NASA/ESA Hubble Space Telescope obtained from the Space Telescope Science Institute, which is operated by the Association of Universities for Research in Astronomy, Inc., under NASA contract NAS 5–26555. These observations are associated with program 8301.
    This research made use of the following software: astropy \citep{2013A&A...558A..33A,2018AJ....156..123A},  
          APLpy \citep{2012ascl.soft08017R}, Numpy \citep{2011CSE....13b..22V,2020Natur.585..357H}, Scipy \citep{jones2001scipy}, CIAO \citep{2006SPIE.6270E..1VF}, XSPEC \citep{1996ASPC..101...17A}, Proffit \citep{2011A&A...526A..79E}, AIPS \citep{van1996aips}, CASA \citep{2007ASPC..376..127M}, BCES \citep{1996ApJ...470..706A}.  
\end{acknowledgements}

%

\newcommand{\noopsort}[1]{} \newcommand{\printfirst}[2]{#1}
  \newcommand{\singleletter}[1]{#1} \newcommand{\switchargs}[2]{#2#1}

%
\appendix{}
\section{Alternative test on the X-ray spectrum of the BCG}\label{altbcg}
In this Appendix we show the results of using the blank-sky event file to model the background for the spectral analysis of the X-ray emission of the BCG in ZwCl~235. We extracted the source and background spectrum from a circle of radius 3$''$ centered on the BCG from the \textit{Chandra} observation and the blank-sky event file, respectively. The use of a local background in \cref{bcg} allowed to remove the contribution of the thermal ambient gas from the BCG spectrum. In order to account for this component when using the blank-sky event file, we included an additional \texttt{apec} component in our fit which is meant to describe the ICM projected in front of the central source. Following the study performed in \cref{bcg}, we considered both a non-thermal and a thermal origin for the X-ray emission. \\ \indent In the first place, we tried to fit the spectrum with a \texttt{tbabs$\ast$(po+apec)} model. The power-law spectral index, when left free to vary, assumes unphysical values; thus, we fixed it to 1.9 (as done in \cref{bcg}). From the best-fit values reported in Tab. \ref{tab:blanktbabsapec}, we note that the power-law normalization is not constrained, resulting in an upper limit similar to that reported in Tab. \ref{tab:bcgspectrum}. Additionally, there are positive residuals around 0.8-1.0 keV similar to those found when using a local background. As a consequence, the residuals were not caused by a poor choice of background, but are rather an intrinsic source component. \\ \indent  In the second place, the spectrum was fit in the 0.5-7 keV band with a \texttt{tbabs$\ast$(apec+apec)}. The temperatures and normalizations of the two \texttt{apec} component were left free to vary, while we linked the abundances of the two \texttt{apec} together due to the low statistics. The results are reported in Tab. \ref{tab:blanktbabsapec}. We identify the component with $kT=2.53^{+0.54}_{-0.32}$ keV as the one describing the ambient gas, while the one with $kT=1.10^{+0.27}_{-0.16}$ keV as the one describing the thermal gas in the inner 5 kpc. The latter temperature is slightly lower than that presented in \cref{bcg}, but fully consistent within 1$\sigma$ errors. From the normalization of the second \texttt{apec} and using a volume $V = 4/3\pi r^{3}$ with $r=4.7$ kpc (the first \texttt{apec} component removes the contribution of the ICM projected along the line of sight), it is straightforward to obtain the electron density $n_{e}=0.06^{+0.02}_{-0.04}$ cm$^{-3}$ (Eq. \ref{normne}). This corresponds to a pressure p$=1.68^{+1.6}_{-0.8}\times10^{-10}$ erg cm$^{-3}$ (Eq. \ref{press}), an entropy K$=7.8^{+4.3}_{-2.2}$ keV cm$^{2}$ (Eq. \ref{entro}), and a cooling time $t_{\text{cool}}=355^{+250}_{-150}$ Myr (Eq. \ref{coolt}), which are all consistent with the values presented in \cref{bcg} within errors.

\begin{table*}
	\centering
	\setlength\tabcolsep{5pt}
	\renewcommand{\arraystretch}{1.5}
	\caption{Spectral analysis of the X-ray spectrum of the BCG using the blank-sky event file as background. 
	}
	\label{tab:blanktbabsapec}
	\begin{tabular}{l|ccc|cc|ccc}
		\hline
		   Model &$kT$ &$Z$&$norm$& $\Gamma$ & $norm$ &$kT$ &$norm$ & $C/d.o.f.$ \\
		    &[keV] &[Z$_{\odot}$]& [ph keV$^{-1}$cm$^{-2}$s$^{-1}$]&  & [ph keV$^{-1}$cm$^{-2}$s$^{-1}$] &[keV] & [ph keV$^{-1}$cm$^{-2}$s$^{-1}$] &  \\
		\hline
		    Non-thermal & $1.83^{+0.17}_{-0.16}$ & $0.95^{+0.28}_{-0.22}$& $1.75^{+0.23}_{-0.22}\times10^{-4}$ & 1.9 &  $<7.09\times10^{-6}$& (...) & (...) & 165/148\\
		    Thermal &  $2.53^{+0.54}_{-0.32}$ & $1.44^{+0.73}_{-0.55}$& $1.19^{+0.23}_{-0.24}\times10^{-4}$ & (...) & (...) &$1.10^{+0.27}_{-0.16}$ & $1.28^{+1.01}_{-0.91}\times10^{-5}$ &158/148\\
		  \hline
	\end{tabular}
	\tablefoot{(1) model used to fit the spectrum; (2-3-4) temperature, abundance and normalization of the \texttt{apec} component that accounts for the ambient gas; (5-6) spectral index and normalization of the power-law; (7-8) temperature and normalization of the thermal model; (9) $C$-statistics/degrees of freedom.}
\end{table*}

\section{Volume of the filament}
\label{appendix:volume}
In this section we detail the method we employed to measure the volume occupied by the two thermal components found in the bright X-ray filament (see \cref{brightf}). In the following, $V_{\text{tot}}$ refers to the total volume of the region used for spectral extraction, while $V_{1}$ and $V_{2}$ refer to the volumes occupied by the cold and hot phases, respectively. The region used for extracting the X-ray properties of the filament, shown in Fig. \ref{figband}, consists in an annular sector extending from $r_{\text{in}}=3''$ to $r_{\text{out}}=13''$ from the center and with an opening angle of 75$^{\circ}$. Using a projected thermal model, the normalization returned by the fit is referred to a \textit{projected volume} given by the intersection of the spherical shell with a cylinder defined by the width of the annulus and infinite height. Therefore, the total volume can be defined as:
\begin{equation}
\label{vtot}
    V_{\text{tot}} = \frac{4}{3}\,\pi\,(r_{\text{out}}^{2} - r_{\text{in}}^{2})^{3/2} = V_{1} + V_{2}
\end{equation}
Since the fitting has been performed using a 2T model, it is necessary to know the relative contribution of the two thermal components to the total volume, known as the filling factor $f = V_{1}/V_{2}$. Following \citet{2011ApJ...732...13G}, assuming that the two spectral phases are in pressure equilibrium in the same volume the filling factor can be estimated as:
\begin{equation}
\label{fill}
    f = V_{1}/V_{2} = \frac{norm_{1}}{norm_{2}}\,\Bigg(\frac{kT_{1}}{kT_{2}}\Bigg)^{2}
\end{equation}
where $norm_{i}$ and $kT_{i}$ are the normalization and temperature of the i-th component. By combining Eq. \ref{vtot} and \ref{fill} we estimated the volume of the cold gas component reported in \cref{originfil} using the following expression:
\begin{equation}
    V_{1} = \Bigg(\frac{f}{1+f}\Bigg)\, V_{\text{tot}}
\end{equation}
For $norm_{1} = 5.7\times10^{-5}$, $norm_{2} = 2.7\times10^{-4}$, $kT_{1}=1.31$ keV and $kT_{2}=2.66$ keV we measure a filling factor $f=0.051$ and a volume $V_{1} = 3.2\times10^{4}$ kpc$^{3}$. 
\\ \indent  For comparison, by approximating the volume of the filament as a cylinder with radius 5 kpc and length 17 kpc (determined from the lowest contour shown in IM2 of Fig. \ref{figband}), we obtain a very similar volume of $V_{1} = 3.4\times10^{4}$ kpc$^{3}$.
\end{document}